\documentstyle[epsf]{mn}
\begin{document}
\title[Dark and visible matter in LSB galaxies]{The dark and
  visible matter content of low surface brightness disk galaxies}
\author[W.J.G. de Blok and S.S. McGaugh] {W.J.G.~de~Blok$^1$ and S.S.
  McGaugh$^2$\\ $^{1}$Kapteyn Astronomical Institute, P.O.~Box 800,
  9700 AV Groningen, The Netherlands\\ $^{2}$Department of Terrestrial
  Magnetism, Carnegie Institution of Washington, 5241 Broad Branch
  Road NW, Washington, DC 20015, USA}
 \date{received: ; accepted: }
\maketitle
\begin{abstract}
  We present mass models of a sample of 19 low surface brightness
  (LSB) galaxies and compare the properties of their constituent mass
  components with those of a sample of high surface brightness (HSB)
  galaxies.  We find that LSB galaxies are dark matter dominated.
  Their halo parameters are only slightly affected by
  assumptions on stellar mass-to-light ratios.  Comparing LSB and
  HSB galaxies we find that mass models derived using the maximum disk
  hypothesis result in the disks of LSB galaxies having systematically
  higher stellar mass-to-light ratios than HSB galaxies of similar
  rotation velocity. This is inconsistent with all other available
  evidence on the evolution of LSB galaxies. We argue therefore that
  the maximum disk hypothesis does not provide a representative
  description of the LSB galaxies and their evolution.  Mass models
  with stellar mass-to-light ratios determined by the colors and
  stellar velocity dispersions of galactic disks imply that LSB
  galaxies have dark matter halos that are more extended and less
  dense than those of HSB galaxies. Surface brightness is thus related
  to the halo properties.  LSB galaxies are slowly evolving, low
  density and dark matter dominated galaxies.
\end{abstract}

\def\HI{H{\sc i}\ }

\section{Introduction}

The discrepancy between the amount of matter implied by the \HI
rotation curves of spiral galaxies and the amount of matter actually
observed (in the form of stars and gas) is usually interpreted as dark
matter (DM) halos that surround the directly observable parts of
galaxies, although alternative theories of modified Newtonian dynamics
(Milgrom 1983) also provide an efficient description of the
observed rotation curves (Sanders 1996).  Early studies by
e.g. Kalnajs (1983) claimed that, based on the optical rotation curves
that were used, there was no need to invoke DM within the optical
radius for some galaxies.  \HI observations, however, showed a
dramatically different picture (Bosma 1978, Begeman 1987).  Substantial
amounts of DM were needed to describe the observed flat rotation
curves outside the optical disk.

Measurements of the distribution of the DM must be extracted from the
rotation curve.  This is usually done by computing the rotation curves
of the visible matter and subtracting these from the observed total
rotation curve.  The residuals show the dynamical signature of DM.
This procedure typically has the following free parameters: a length scale
and density for the halo and a mass-to-light ratio
$(M/L)_{\star}$ of the stellar component.  A major problem is that the
value of $(M/L)_{\star}$ is not known {\it a priori}, and may
differ from galaxy to galaxy.  Many different values of
$(M/L)_{\star}$, and therefore many combinations of halo parameters,
yield equally good descriptions of the data (van Albada et al.\
1985).  It is for many galaxies possible to make a good fit to the
rotation curve by completely ignoring the contribution of the stellar
disk, even though it obviously plays an important role in the inner
parts of many galaxies.  Without additional knowledge to
constrain $(M/L)_{\star}$ (e.g.\ information about vertical stellar
velocity dispersion or derived from stellar population synthesis
models) it is not possible to unambiguously determine its value.
Goodness-of-fit estimators are sensitive to small variations and
uncertainties in the data (even when the data is of superior quality)
(Lake \& Feinswog 1989) and are thus less suited for simultaneously
determining {\it all} free parameters in a fit. The mathematically ``best'' fit
is therefore not always the physically most meaningful one.

Using the observation that the rotation curves derived for the stellar
components of HSB galaxies can usually be scaled so that they almost
completely describe the observed total rotation curves in the optical
disks of these HSB galaxies (e.g. Kent 1986), van Albada \& Sancisi (1986)
introduced the maximum-disk hypothesis. This hypothesis reduces the
number of free parameters in rotation curve fits by maximizing the
contribution of the stellar disk [i.e. $(M/L)_{\star}$] and thus
minimizing the amount of DM invoked to explain the observed rotation
curves.

It is still unclear whether the maximum disk hypothesis is a realistic
one. For some HSB galaxies stellar population synthesis models
yield values of $(M/L)_{\star}$ that are consistent with
the maximum disk measurements (van Albada \& Sancisi 1986).  The
presence of disk features such as spiral arms or bars argues for a
high $(M/L)_{\star}$ (Athanassoula, Bosma \& Papaioannou 1987).
Kuijken \& Gilmore (1989) showed, however, that there is only enough
matter in the disk of our Galaxy for it to be about half-maximum disk.
Measurements of the stellar velocity dispersions in HSB galaxies
(Bottema 1995) show that maximum disk may overestimate the amount of
matter present in the disk by some 60 per cent.

As the maximum disk hypothesis gives for each galaxy the {\it maximum}
possible $(M/L)_{\star}$, any conclusions about structural parameters
of halos are only {``lower limits.''}  Nevertheless, attempts have
been made to find systematic relations between the observable
properties of spiral galaxies and their dark halo properties.

A general conclusion is that the importance of DM increases towards
later types (Persic \& Salluci 1991). Coupled with the finding that
the surface brightness also decreases towards later types (de Jong
1995), this means that low surface brightness (LSB) galaxies are a
prime example of galaxies that are extremely interesting for deriving
structural properties of DM halos.

These galaxies are very late-type spirals with central surface
brightnesses much lower than those of ``classical'' late-type HSB
galaxies (de Blok, McGaugh \& van der Hulst 1996 [hereafter BMH96],
Zwaan et al.  1995, van der Hulst et al.\ 1993; see also the reviews
by Impy \& Bothun 1997 and Bothun, Impey \& McGaugh 1997).  Their blue
colors (de Blok, van der Hulst \& Bothun 1995, McGaugh \& Bothun
1994, R\"onnback \& Bergvall 1994), low star formation rates (McGaugh
1992, van der Hulst et al.\ 1993, van den Hoek et al.\ 1997), low oxygen
abundances (McGaugh 1994, R\" onnback \& Bergvall 1995) and high
gas-fractions (BMH96, McGaugh \& de Blok 1997), all give rise to the
picture that these galaxies are slowly evolving and still in an early
evolutionary state.

Recently a number of rotation curves of LSB galaxies have been measured
(BMH96).  A preliminary investigation showed a systematic
trend of increasing total mass-to-light ratio within a number of scale
lengths with decreasing surface brightness, from values of $\sim 1$
for HSB galaxies to $\sim 10$ for the dimmest measured LSB galaxies
(Fig.\ 13 in BMH96), consistent with predictions made using the
Tully-Fisher relation (Zwaan et al.\ 1995).  The increasing importance
of DM makes an investigation into the halo parameters extra
interesting as the parameters will depend less strongly on assumed
values of $(M/L)_{\star}$, and one can therefore be quite sure that
the derived parameters are indeed close to the intrinsic parameters of
the halo.

A comparison of a HSB and a LSB galaxy at identical positions on the
Tully-Fisher relation (i.e. at fixed $V_{\rm max}$) (de Blok \&
McGaugh 1996) showed that LSB galaxy UGC 128 is almost a factor 10
less dense than the HSB galaxy NGC 2403 (analogous with their low
stellar and \HI surface densities).  UGC 128 also has a factor of 7
higher total $M/L$ (measured within a fixed number of scale lengths)
and must therefore be more massive than HSB galaxy NGC 2403 (which has
the same luminosity).

In this paper we will discuss disk-halo decompositions of the sample
of LSB galaxies from BMH96 and compare the structural parameters of
the dark and luminous mass components of HSB and LSB galaxies.  In
Section 2 we describe the sample; Section 3 covers various aspects of
the effects on resolution and beam-smearing on the rotation curves.
In Section 4 the actual decompositions are presented, while in Section
5 we discuss the results and compare them with what is known about
other galaxies.  In Section 6 the systematic relations between the
masses of the different components (gas, stellar, dark) are discussed,
while in Section 7 we discuss the trends of volume and surface density
as a function of surface brightness.  In Section 8 our main results
are summarized.  In Appendix A a summary of the Bottema disk
(Sect. 4.4) is given.

\section{The LSB Sample}

We refer to BMH96 for an extensive sample description.  We have
supplemented that sample with the rotation curves of an additional 5
LSB galaxies as presented in van der Hulst et al.\ 1993.  The rotation
curves in general have maximum rotation velocities between 50 and 120
km s$^{-1}$, and rise more slowly than HSB curves with similar maximum
velocity.  Only a few of the rotation curves show clear signs of
flattening towards the outermost radii.  Optical properties of the
sample are presented in McGaugh \& Bothun (1994), de Blok et al.\
(1995), and van der Hulst et al.\ (1993).  The properties of the
sample galaxies are summarized in the top panel of Table 1.  We adopt
a Hubble constant of 75 km s$^{-1}$ Mpc$^{-1}$.

\begin{table*}
\begin{minipage}{110 mm}
\caption[]{Properties of LSB and HSB samples}
\begin{tabular}{lrcccccccc}
  
  \hline Name& $D$&$\mu_0(B)$&$h$& $B-V$& $M_{\rm abs}(B)$&
  $R_{25}$&$R_{\rm max}$& $V_{\rm max}$\\ & (Mpc)&(mag/$\sq
  "$)&(kpc)&(mag)&(mag)& (kpc)&(kpc)&(km s${^{-1}}$)\\ \hline
  \multicolumn{3}{l}{{\bf LSB sample} (this work)}&&&&&&&\\ \hline
  F561-1 & 63 & 23.3 & 3.6 & 0.55 & -17.8 & 6.4 & 10.1 & 52\\ F563-1 &
  45 & 23.6 & 2.8 & 0.64 & -17.3 & 5.0 & 17.7 & 111\\ F563-V1 & 51 &
  24.3 & 2.4 & 0.59 & -16.3 & 3.2 & 7.4 & 30\\ F563-V2 & 61 & 22.1 &
  2.1 & 0.51 & -18.2 & 6.8 & 9.2 & 111\\ F564-V3 & 6 & 24.1 & 0.4 &
  0.56 & -11.1 & 0.3&\llap{$\sim$}2&\llap{$\sim$}40\\ F565-V2 & 48 &
  24.7 & 2.7 & 0.51 & -15.4 & 2.3 & 8.4 & 51\\ F567-2 & 75 & 24.4 &
  5.7 & 0.61 & -17.4 & 3.2 & 11.3 & 64\\ F568-1 & 85& 23.8 & 5.3 &
  0.58 & -18.1 & 7.4 & 14.9 & 119\\ F568-3 & 77 & 23.1 & 4.0 & 0.61
  &-18.3 & 7.9 & 16.5 & 120\\ F568-V1 & 80 & 23.3 & 3.2 & 0.57 & -17.9
  & 6.2 & 19.0 & 124\\ F571-8 & 48 & 23.9\rlap{$^{a,c}$} &
  5.2\rlap{$^c$} & * & -17.6\rlap{$^a$} & 7.7 & 15.6 & 133\\ F571-V1 &
  79 & 24.0 & 3.2 & 0.55 & -17.0 & 3.8 & 14.6 & 73\\ F571-V2 & 16 & *
  & * & * & * & * & 3.7 & 45\\ F574-1 & 96 & 23.3\rlap{$^a$} & 4.3 & *
  & -18.4\rlap{$^a$} & 17.7 & 15.4 & 100\\ F574-2 & 88 & 24.4 & 6.0 &
  0.59 & -17.6 & 3.8 & 10.7 & 40\\ F577-V1 & 80 & 24.0 & 4.3 & 0.40 &
  -18.2 & 6.6 & 8.9 & 30\\ F579-V1 & 85 & 22.8\rlap{$^a$} & 5.1 & * &
  -18.8\rlap{$^a$} & 11.1 & 17.3 & 100\\ F583-1 & 32 & 24.1 & 1.6 & *
  & -16.5 & 3.3 & 14.6 & 85\\ F583-4 & 49 & 23.8\rlap{$^a$} & 2.7 & *
  & -16.9\rlap{$^a$} & 6.4 & 10.0 & 67\\ U0128 & 60 & 24.2 & 6.8 &
  0.60 & -18.8 & 9.0 & 42.3 & 131\\ U1230 & 51 & 23.3 & 4.5 & 0.47 &
  -18.3 & 5.2 & 34.7 & 102\\ U5005 & 52 & 23.8\rlap{$^a$} & 4.4 & * &
  -17.8\rlap{$^a$} & 10.4 & 27.8 & 99\\ U5750 & 56 & 23.5\rlap{$^a$} &
  5.6 & * & -18.7\rlap{$^a$} & 9.5 & 21.8 & 75\\ U5999 & 45 &
  23.5\rlap{$^a$} & 4.3 & * & -17.6\rlap{$^a$} & 9.8 & 15.3 & 155 \\ 

\hline 
\multicolumn{3}{l}{{\bf ``HSB'' sample} (Broeils 1992)}&&&&&&\\ 
\hline 

DDO154 & 4 & 23.2 & 0.5 & 0.29 & -13.8 & 1.0 & 7.6 & 48 \\ DDO168 &
3.5 & 23.4 & 0.9 & 0.22 & -15.2 & 1.7 & 3.4 & 55 \\ DDO170 & 12 & * &
1.3 & * & -14.5 & 2.5 & 9.6 & 66 \\ N55 & 1.6 & 21.3 & 1.6 & 0.38 &
-18.6 & 8.9 & 10.2& 87 \\ N247 & 2.5 & 23.4 & 2.9 & 0.45 & -18.0 & 7.3
& 9.9 & 108 \\ N300 & 1.8 & 22.2 & 2.1 & 0.56 & -17.8 & 5.1 & 10.6& 97
\\ N801 & 79.2& 21.9 & 12 & 0.66 & -21.7 & 36.4& 58.7& 222 \\ N1003 &
11.8& 21.7 & 1.9 & 0.40 & -19.2 & 9.2 & 31.3& 115 \\ N1560 & 3 & 23.2
& 1.3 & 0.43 & -15.9 & 4.3 & 8.3 & 79 \\ N2403 & 3.3 & 21.4\rlap{$^b$}
& 2.1 & 0.38 & -19.3 & 8.4 & 19.5& 136 \\ N2841 & 18 & 21.1\rlap{$^b$}
& 4.6 & 0.80 & -21.7 & 22.6& 81.1& 323 \\ N2903 & 6.4 &
20.5\rlap{$^b$} & 2.0 & 0.59 & -21.0 & 11.7& 24.2& 201 \\ N2998 &
67.4& 20.3 & 5.4 & 0.48 & -21.9 & 28.2& 46.6& 214 \\ N3109 & 1.7 &
23.2 & 1.6 & * & -16.8 & 4.3 & 8.2 & 67 \\ N3198 & 9.4 &
21.6\rlap{$^b$} & 2.6 & 0.46 & -19.4 & 11.4& 29.9& 157 \\ N5033 &
11.9& 23.0\rlap{$^b$} & 5.8 & 0.48 & -20.2 & 18.3& 35.4& 222 \\ N5533
& 55.8& 23.0 & 11.4 & 0.80 & -21.4 & 26.2& 74.4& 273 \\ N5585 & 6.2 &
21.9 & 1.4 & 0.42 & -17.5 & 4.7 & 9.6 & 92 \\ N6503 & 5.9 & 21.9 & 1.7
& 0.56 & -18.7 & 5.4 & 22.2& 121 \\ N6674 & 49.3& 22.5 & 8.3 & 0.61 &
-21.6 & 29.8& 64.5& 266 \\ N7331 & 14.9& 21.5\rlap{$^b$} & 4.5 & 0.70
& -21.4 & 23.4& 36.7& 241 \\ U2259 & 9.8 & 22.3\rlap{$^b$} & 1.3 & * &
-17.0 & 3.7 & 7.6 & 90 \\ U2885 & 78.7& 22.0\rlap{$^b$} & 13 & * &
-22.8 & 62.8& 72.5& 298 \\ \hline
\end{tabular}
$a$: converted from $R$-band measurement\\ $b$: derived from $r$-band
profiles, converted using $B=r+1$ (Kent 1986)\\ $c$: $\mu_0$(face\ on)
of F571-8 was computed assuming $\mu{\rm (face\ on)} = (z_0/h)\mu{\rm
  (edge\ on)}$ (i.e. optically thin). $h{\rm(face\ on)}$ was found to
be $\sim 1.4 h{\rm(edge\ on)}$ following method of Peletier et al.\ 
(1995).\\ 
\end{minipage}
\end{table*}   

As described in BMH96 the rotation curves were derived from major-axis
position-velocity diagrams. Inclinations were determined from a
comparison of optical, H{\sc i} and kinematical inclinations. The
position angle and inclination were both kept fixed during the
derivation of the curves.  Inspection of the position-velocity
diagrams in Fig.\ 2 of BMH96 shows that not all rotation curves are
well suited for a disk-halo analysis. A number of observations clearly
suffer from low resolution or low inclination.

In order to select only those galaxies where a disk-halo decomposition
is a sensible exercise, we restricted the sample by rejecting those
galaxies that did not meet one of the following criteria:
\begin{enumerate}
\item inclination $i > 25^{\circ}$,
\item radius $R > 2$ beams,
\item asymmetries between approaching and receding sides of rotation
  curves $< 10$ km s$^{-1}$.
\end{enumerate}
This resulted in the rejection of seven galaxies. These are listed in
Table 2 with the reason for rejection.

The choice for a 25 degree cut-off may seem a rather liberal choice,
which in a sense we are forced to apply due to the selection biases
against high-inclination LSB galaxies (Davies 1990, Schombert et al.\ 
1992, Dalcanton \& Shectman 1996).  An analysis of the scatter in the
LSB TF relation presented in Zwaan et al. (1995) does show however
that this scatter does not increase significantly when increasingly
lower-inclination galaxies are included until a value of 25 degrees.

The effects of lower resolution are more difficult to model, and will
be described extensively in the next section.

\begin{table}
\begin{minipage}{60 mm}
\caption[]{Rejected galaxies}
\begin{tabular}{ll}
  \hline Name & Limitation\\ \hline F561-1 & $R < 2$ beams; $i <
  25^{\circ}$\\ F563-V1 & $R < 2$ beams; gross asymmetry \\ F564-V3 &
  $R < 2$ beams; low S/N\\ F567-2 & gross asymmetry; $i <
  25^{\circ}$\\ F574-2 & $R < 2$ beams; $V(R)$ steeply rising\\ 
  F577-V1 & $R < 2$ beams; gross asymmetry\\ F579-V1 & gross
  asymmetry\\ \hline
\end{tabular}
\end{minipage}
\end{table}

\section{Beam-smearing}

Early high-resolution studies of spiral galaxy kinematics using
synthesis radio telescopes did mostly target galaxies with large
angular sizes (e.g.\ Bosma 1978, Begeman 1987, Broeils 1992).  The
rotation curves of these galaxies are in general well-determined.
Unfortunately there are only a limited number of galaxies in the sky
that have suitably large angular sizes for such detailed
investigations.  Galaxies with smaller apparent sizes (because of
smaller intrinsic sizes and/or larger distances) can only be observed
at a smaller linear resolution.

A side-effect of this lower resolution is that the line profiles will
be artificially broadened and any sudden change in the velocity
gradient within the beam will be smoothed out.  The observed change in
the velocity gradient will thus appear to be smaller and the rotation
curve will appear to rise less steeply. This effect is called ``beam
smearing.''

Bosma (1978) investigated the behavior of a steeply rising model
rotation curve at different resolutions.  He showed that if the ratio
between the Holmberg radius of a galaxy and the half power beam width
is smaller than $\sim 7$ to $\sim 10$, beam smearing can have serious
effects on the steep inner parts of the curve.  Rubin et al.\ (1989)
make the point that if one wants to decompose the mass distribution of
a galaxy into its disk and halo components one needs rotation curves
of high accuracy in their inner portions, as ``the maximum mass which
can reside in the disk is constrained principally by the inner rise of
the rotation curve.''  This statement is based on investigations into
a sample of primarily steeply rising rotation curves, but
remains true also for slowly rising rotation curves, although, as we
will show, to a much lesser extent. 

The rotation curves of LSB galaxies presented in BMH96 are not
high-resolution rotation curves and beam smearing is a potential
problem. The decrease in the slope of the rotation curves would
disguise potentially steep rotation curves as gently rising solid-body
curves.  Here we will present some arguments why this effect is not
significant in the BMH96 data, implying that the rotation curves of
LSB galaxies are truly slowly rising.

\subsection{Evidence from BMH96}

The most direct evidence why beam smearing effects do not dominate the
data from BMH96 comes from that data itself.  If the data were
severely affected by resolution effects, one would not expect to
observe steeply rising rotation curves.

\begin{figure*}
\epsfxsize=\hsize
\hfill\epsfbox{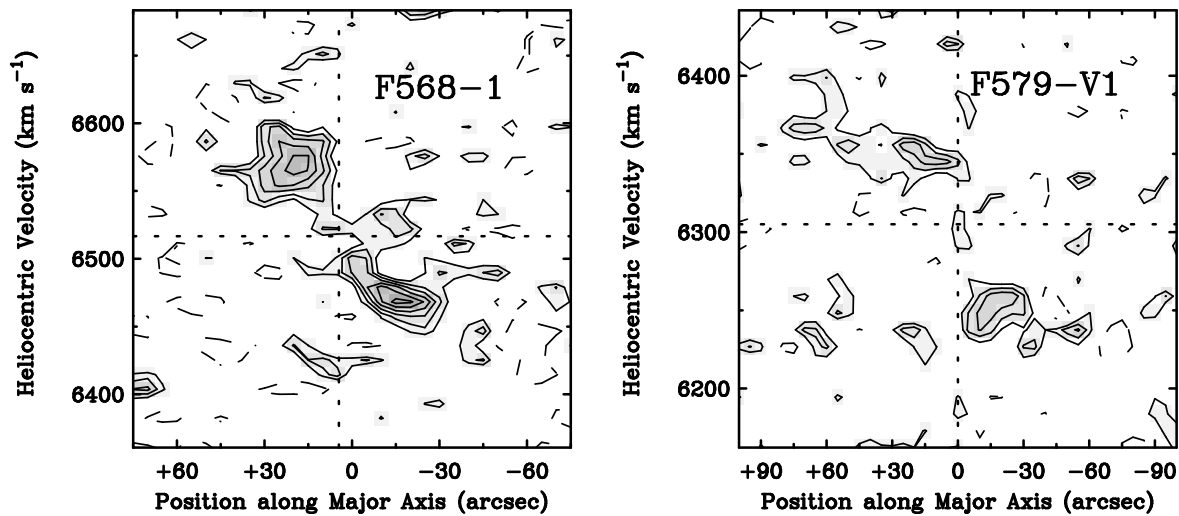}\hfill
\caption{Major-axis position velocity diagrams of LSB galaxies F568-1
and F579-V1. Both galaxies are at approximately the same distance,
have identical inclinations and were observed with the same telescope
with identical beam sizes. It is clear that the rotation velocity in
the inner parts of F579-V1 increases more rapidly than in F568-1. If
these observations suffered from heavy beam-smearing this difference
could never have been observed. Contour levels are -2$\sigma$ (dashed
contours) and 2,4,6,.. $\sigma$ contours (full contours).}
\label{579v1_5681_beam}
\end{figure*}

Figure \ref{579v1_5681_beam} shows the major-axis position-velocity
diagrams of LSB galaxies F579-V1 and F568-1.  Both galaxies are at
similar distances (77 {\it vs} 85 Mpc, a difference of less than 10
per cent); both were observed with the same telescope with identical
beam sizes; both have an inclination of 26 degrees, a comparable \HI\
distribution and almost identical scale lengths.  The effects of beam
smearing are effectively {identical} in both galaxies.

It is clear that the rotation curve of F579-V1 rises much more steeply
than that of F568-1. The fact that we {\it do} observe a steeply
rising rotation curve in F579-V1, shows that the approximately solid
body rotation observed in F568-1 is not caused by beam-smearing.  This
makes it unlikely that the solid-body rotation observed in other LSB
galaxies from BMH96 is caused by observational effects. It should be
noted that the solid-body rotation is also seen in those LSB galaxies
observed at higher resolutions (e.g. F563-1 and F583-1) and is in
general found in other high resolution observations of very late-type
galaxies (see e.g. IC 2574 in Martimbeau, Carignan \& Roy 1994).

\subsection{Modeling beam smearing}

Another way to quantify the effects of beam smearing is to construct
model galaxies with known properties and ``observe'' these at
increasingly lower resolutions.  As beam smearing affects the full
two-dimensional velocity field of a galaxy, it is necessary to
construct complete model data cubes, smooth these to lower resolutions
using a Gaussian beam and then construct velocity fields and derive
rotation curves.

\subsubsection{Constructing models}

We constructed a number of model data cubes using the task {\sc
  galmod} in the Groningen Image Processing System {\sc gipsy}. This
program distributes ``\HI clouds'' in a specified data cube using an
input radial \HI distribution and rotation curve as distribution
functions.  The galaxy can be given any inclination or position angle.
This model can then be ``observed'' at any desired spatial and
velocity resolution.  For simplicity we adopted a uniform, constant
density \HI distribution for all models. We could have used any \HI
distribution, but decided to opt for the simplest distribution to
isolate the effects of beam smearing.

The magnitude of the beam smearing effects also depends on
inclination. At high inclinations the beam will cover a larger part of
the velocity field off the major axis than at lower inclination, thus
``diluting'' the major axis rotation velocity.  For a fixed beam size
the effects of beam smearing will still depend on inclination, with
the worst effects occuring in the highest-inclination galaxies.

\begin{figure*}
\epsfxsize=\hsize
\hfil\epsfbox{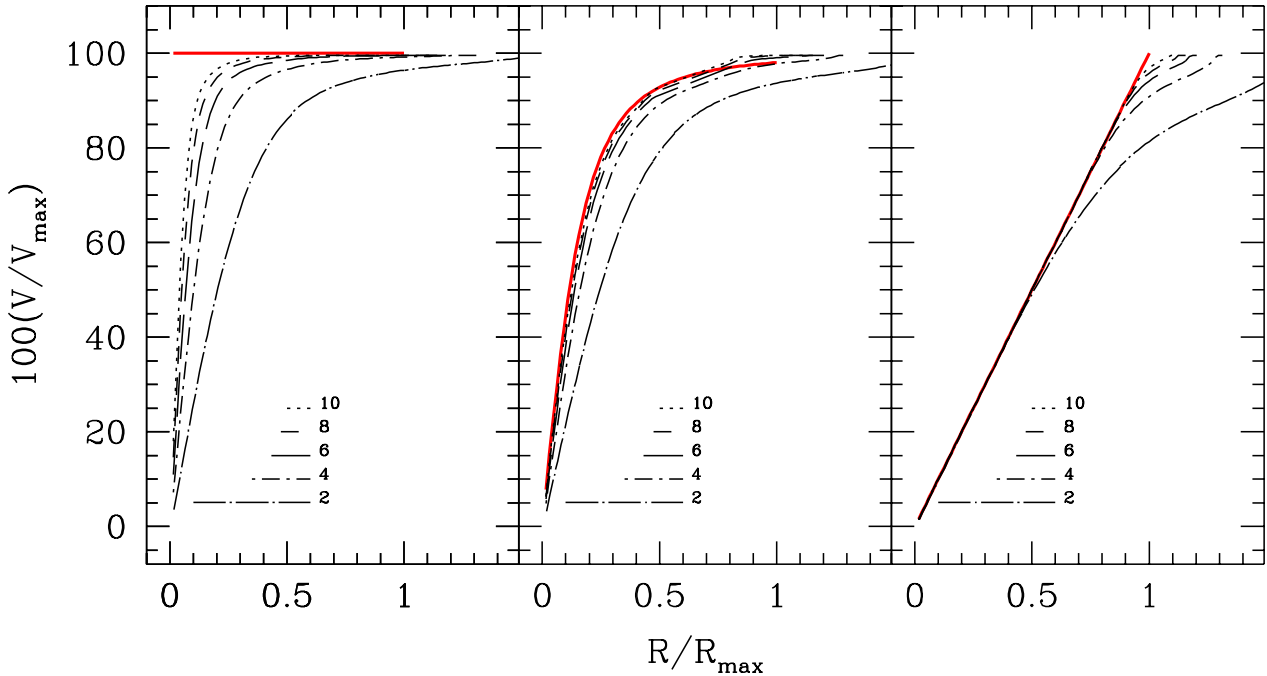}\hfil
\caption{Model rotation curves illustrating the effects of beam-smearing.
The heavy lines show the input rotation curves: flat (left panel),
late-type-like (center), and solid-body (right). The beam-smeared
versions of these curves are shown as the dashed and dotted
lines. These lower resolution curves were derived using the full
3-dimensional data cubes. The resolution expressed in beams is 10
(dotted), 8 (short dash), 6 (long dash), 4 (short dash-dot) and 2
(long dash-dot). It is clear that the effects of beam-smearing are
most severe in the flat case (left panel), while the effects of
beam-smearing are only small in the inner parts of the late-type and
solid-body curves. In the latter two cases the shape of the curves do
not change dramatically down to a resolution of 2 beams. The small
bump visible in the FE-curves at $R \simeq 1$ is caused by the limited
velocity resolution used in the models, in order to mimic the BMH96
observations. }
\label{beamsmearcurves}
\end{figure*}

In order to mimic the BMH96 observations, we chose an inclination of
40 degrees for our model galaxies, equal to the average inclination of
the 12 ``F''-galaxies in BMH96. To cover the full range of known
rotation curve shapes we adopted three different input rotation
curves. These are shown in Fig. \ref{beamsmearcurves}. Radii and
velocities are all expressed as fractions of the maximum radius and
velocity of the input models $R_{\max}$ and $V_{\rm max}$.  These were
identical for all three models.

The first model has a constant rotation velocity at all radii, and is
the perfect example of a flat rotation curve.  This is an extreme, but
not unrealistic representation of the steeply rising rotation curves
found in early-type HSB galaxies (e.g. UGC 2885 or NGC 2841 in the
compilation by Rhee 1996).

The second curve is a simplified version of that presented in Fall \&
Efstathiou (1980): \[ V(r) = V_{\rm max} \sqrt {{r^2}\over{r^2+d^2}}.
\] The constant $d$ was given a value of 0.2$R_{\rm max}$. This model
rotation curve (which we will refer to as the ``FE curve'') resembles
that of a normal late-type galaxy (see e.g.  the high resolution
rotation curves of NGC 1560, NGC 3198 or NGC 2403 in Rhee 1996).

The third curve is another extreme case: pure solid body rotation with
a still increasing rotation velocity at $R_{\rm max}$. This solid-body
rotation {\it is} found in dwarf galaxies (see e.g. the well defined
high resolution curve of IC 2574 in Martimbeau et al.\ 1994 or NGC
3109 from Rhee 1996).

The full three-dimensional high-resolution model data cubes
constructed using these curves, were each smoothed with Gaussian beams
with FWHM sizes of $(1/10, 1/8, 1/6, 1/4, 1/2) \times R_{\rm max}$.
This obviously resulted for each of the three cases in data cubes
where the {\it radius} of the disk measured 10, 8, 6, 4 and 2 beams
respectively.  Velocity fields were constructed from these cubes using
the {\sc gipsy} task {\sc moments}.  Analogous to the procedure
outlined in BMH96, the rotation curves of each of the smoothed data
cubes were determined from major-axis position-velocity diagrams.

As noted before, all models were computed assuming model galaxy
inclinations of 40 degrees.  The original and smoothed 
de-projected curves are presented in Fig.~\ref{beamsmearcurves}.  A
first glance shows that indeed the effects of beam smearing are most
severe for the flat rotation curve which has the steepest inner
``slope.'' The beam smearing effects are negligible in the solid body
curve, where the only effect is an apparent flattening at the
outermost radii. The FE curve retains its shape and slope down to
a resolution of 4 beams. The input and output curves only start to
differ significantly at resolutions of 2 beams. 

\subsubsection{Comparison with observations}

\begin{figure}
\epsfxsize=\hsize
\hfil\epsfbox{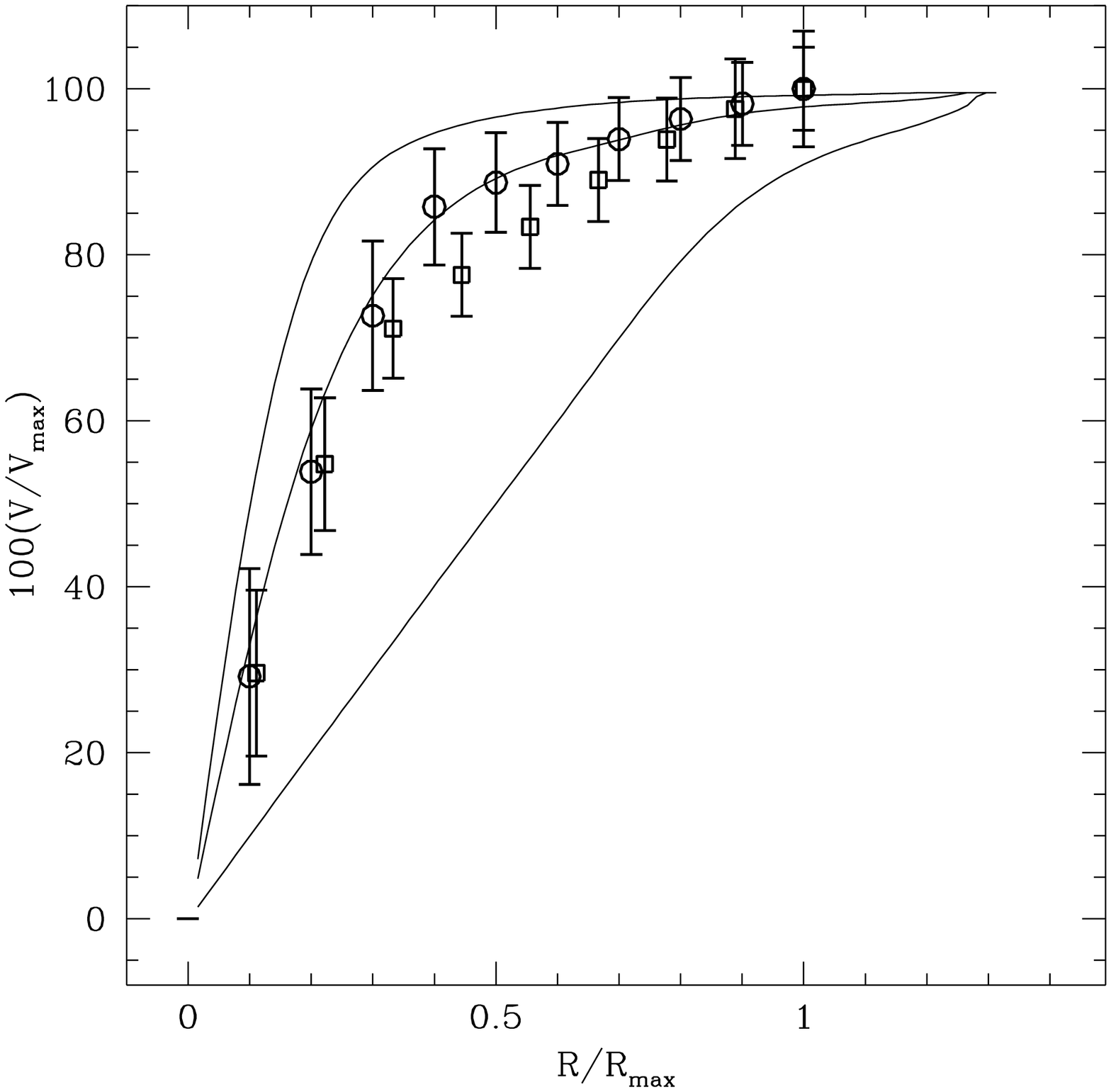}\hfil
\caption{Comparison of the model rotation curves with observed curves.
The three light lines show, from top to bottom, the rotation curves of
the flat model, the FE model and the solid-body model at a resolution
of 4 beams. Super imposed are the rotation curves of
F568-V1 (squares) and F568-3 (circles). The latter curves are consistent
with the FE-model, although the shapes suggest that a slightly steeper
solid-body model is also possible. The flat model can be excluded. }
\label{compareres}
\end{figure}

We now compare the smoothed models with some of the observations from
BMH96.  We will use the rotation curves of F568-3 and F568-V1. These
galaxies both have inclinations of 40 degrees and are thus directly
comparable with the model curves. Both curves have a resolution of
slightly less than 4 beams and should thus be compared to the 4-beam
models, as is done in Fig.~\ref{compareres}.  The observed curves were
scaled using the $R_{\rm max}$ and $V_{\rm max}$ given in BMH96.

It is clear that the observed curves are best matched by the FE curve,
although a completely solid body curve cannot be ruled out at 4 beams
resolution. The only way the observations can be reconciled with a
rotation curves that are flat down to a small radius, is at a
resolution of 2 beams, that is, a factor of 2 worse than the present
observations. At a resolution of 4 beams a flat curve can be ruled out.

The question that remains to be answered then, is how likely it is
that LSB galaxies have FE-like rotation curves.  First, as is shown by
the studies of Casertano \& van Gorkom (1989) and Broeils (1992) of
high resolution \HI rotation curves, {\it only} those galaxies with
maximum rotation velocities of more than 180 km s$^{-1}$ ($M_B \leq
-19$) have steep inner rotation curves. The steep inner curves are
usually interpreted as the dynamical signature of a strong bulge
component.  Galaxies that rotate slower have gradually rising rotation
curves, resembling the FE model curve.  The LSB galaxies from BMH96
have maximum rotation velocities between 50 and 120 km s$^{-1}$,
luminosities $M_B$ between $-16$ and $-18$, and have only the faintest
trace of a central condensation. So, {\it if} LSB galaxies are
late-type galaxies, as all of the available evidence suggests, then it
is not surprising that they have slowly rising rotation curves.

This is supported by work from Coradi \& Capaciollo (1990) who
measured the inner slopes of {\it optical} rotation curves of 139
galaxies of a large range of Hubble types and luminosities. They found
a strong correlation between the magnitude of this slope, Hubble
type and luminosity: 75 per cent of the galaxies in their sample with
luminosities $\-19 < M_B \leq -18$ have slowly rising, non-flat
rotation curves\footnote{Note that the steep slope of the F579-V1
  rotation curve is consistent with these results. F579-V1 is the most
  luminous galaxy in the ``F''-sample from BMH96, and in fact falls in
  the $\-19 < M_B \leq -18$ category.}, while all of the galaxies
fainter than $-18$ have slowly rising rotation curves. The result
holds for the relation with Hubble types: 80 per cent of the galaxies
with Hubble type Sd and later have slowly rising rotation curves.

\subsection{Evidence from NGC1560}

\begin{figure*}
\epsfxsize=\hsize
\hfil\epsfbox{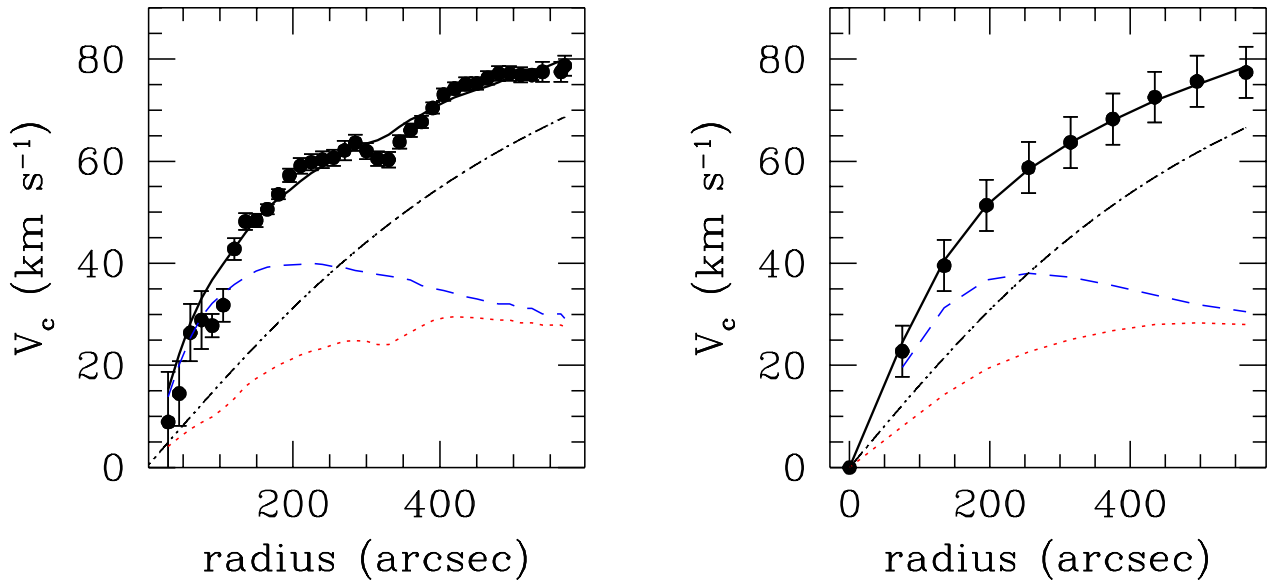}\hfil
\caption{High and low resolution rotation curves 
of NGC 1560. The left panel shows the original high resolution
observation from Broeils (1992). The curves measures 38 beams, and is
well-resolved. The right panel shows the same data but smoothed to a
resolution of only 3 beams. It is striking that the overall shape and
amplitude of the curves have remained virtually unchanged. This is
discussed more fully in Section 3.3. The dashed and dotted lines show
the maximum disk decompositions for both cases. The dotted lines
represent the rotation curve of the gas component, the dashed lines
the rotation curve of the maximum stellar disk. The dash-dotted lines
are the rotation curves of the halos as derived from this
decomposition. There is almost no difference between the low- and
high-resolution halo parameters.
This is discussed more fully in Section 5.1. The
fitting parameters are given in Table 6.}
\label{smoreg1560}
\end{figure*}

As LSB galaxies are late-type galaxies, a third possibility to
investigate the effects of beam-smearing is by taking a high-resolution
observation of a late-type galaxy and smooth it to lower resolution.
We have selected NGC 1560 for this purpose. This galaxy is a LSB dwarf
galaxy ($\mu_0(B) = 23.2\ {\rm mag\ arcsec}^{-2}$, for other
properties see Table 1), for which a high-resolution rotation curve is
available (Broeils 1992). This curve is shown in the left panel of
Fig. \ref{smoreg1560}. The beam size for this WSRT observation was
$13'' \times 14''$, and the extent of the rotation curve is 38 beams.
Combined with the small distance of only 3 Mpc, we can be that sure
beam-smearing is of no importance in this observation.

We have used one-dimensional smoothing with a Gaussian filter to bring
the total rotation curve and those of the stars and gas to lower
resolutions. As the effects of one-dimensional smoothing of the
rotation curve are less severe than full two-dimensional smoothing of
the velocity field, we have smoothed the curves to a resolution of 3
beams. We found by trial and error that this corresponds most closely
to a full two-dimensional smoothing with a resolution of 4 beams.

The smoothed curve is shown in the right panel of
Fig.~\ref{smoreg1560}. The other curves shown in the figure will be
discussed in Section 5.1. A comparison of the high- and low-resolution
curves shows that the only effects this beam smearing exercise has had
on the curves is a loss of detail.  The shapes and amplitudes of the
curves have remained unchanged. This again shows that slowly rising
rotation curves can be studied at much lower resolutions than rapidly
rising ones without a dramatic loss of information.  We conclude that
the rotation curves of the LSB galaxies are intrinsically slowly
rising. Beam smearing effects do not dominate the BMH96 data.

\section{Disk halo decomposition}

We now return to the disk-halo decompositions and subsequent mass
modeling.  Before giving the results of the decompositions, we first
describe the mass components used.

{\bf Stellar disk} For the stellar disk the $B$ and $R$-band
photometry presented in de Blok et al. (1995), and McGaugh \& Bothun
(1994) was used.  The rotation curve of the disk was computed
following Casertano (1983) and Begeman (1987). The disk was assumed to
have a vertical sech-squared distribution with a scale height $z_0 =
h/6$ (van der Kruit \& Searle 1981).  The exact value of $z_0$ does
not influence the amplitude and shape of the rotation curve to any
significant degree.  No bulge component was included as there is no
evidence that bulges contribute significantly to the mass distribution
in most LSB galaxies (see de Blok et al. 1995, McGaugh, Bothun \&
Schombert 1995). Prior to the decomposition the light distribution
curves were resampled to the \HI resolution.

{\bf Isothermal dark halo} We assume a spherical pseudo-isothermal
halo with a density profile
 \begin{equation} \rho(R) = \rho_0 \Bigl[ 1 + \Bigl( {{R}\over{R_C}} \Bigr)^2 \Bigr]^{-1},
 \end{equation}
 where $\rho_0$ is the central density of the halo, and $R_C$ the core
 radius of the halo.  This density profile results in the rotation
 curve
 \begin{equation} 
   V_{\rm halo}(R) = \sqrt{ 4\pi G\rho_0 R_C^2 \Bigl[ 1 -
     {{R_C}\over{R}}\arctan \Bigl( {{R}\over{R_C}} \Bigr) \Bigr] }.
\end{equation}
The asymptotic velocity of the halo, $V_{\infty}$, is given by
\begin{equation} 
  V_{\infty} = \sqrt{ 4 \pi G \rho_0 R_C^2 }.
\end{equation}
To characterize this halo only two out of the three parameters
$(\rho_0, R_C, V_{\infty})$ are needed, as equation (3) determines the
value of the third parameter.  The mass of the halo integrated out to
radius $r$ is
\begin{equation} 
  M_{\rm d}(r) = {{V_{\rm halo}^2 r}\over{G}} = 4 \pi \rho_0 R_C^2 \left[ r - R_C \arctan \left(
      {{r}\over{R_C}}\right) \right].
\end{equation}

Because $(M/L)_{\star}$ is unknown we present disk-halo decompositions
using three different assumptions for $(M/L)_{\star}$:
 \begin{enumerate} 
 \item maximum disk (Sec.\ 4.2),
 \item minimum disk (Sec.\ 4.3), and
 \item ``Bottema disk'' (Sec.\ 4.4).
 \end{enumerate} 

 Each of the rotation curves was fitted using a program which
 determines the best-fitting combination of $R_C$, $V_{\infty}$,
 $(M/L)_{\star}$, using a least squares fitting.  Any combination of
 these three parameters could be fit, or kept fixed at some initial
 value.  The program uses as input the rotation curve of the gas, the
 total measured rotation curve, and the rotation curve of the stellar
 disk computed from the observed light distribution (i.e. with an
 $(M/L)_{\star} = 1$) (see Begeman 1987).

\subsection{HSB and LSB samples}

We will compare the properties of our LSB galaxies with those of
mostly HSB galaxies with well-defined \HI rotation curves.  The latter
have been taken from the compilation by Broeils (1992) and references
therein.  This ``HSB sample'' does not consist of just disks obeying
Freeman's Law (i.e.  galaxies with $\mu_0 \simeq 21.6 B$-mag
arcsec$^{-2}$) (Freeman 1970): a large range of central surface
brightnesses is found, from Freeman disks to galaxies with surface
brightnesses approaching those of the LSB galaxies.  The mean surface
brightness of the HSB sample ($\langle \mu_0 \rangle = 22.2$) is still
$1.5$ magnitudes brighter than that of the LSB sample ($\langle \mu_0
\rangle = 23.7$).  Comparing the two samples gives a good impression
of the change in properties of spiral galaxies over a large range in
surface brightness.  Properties of the HSB sample are given in the
bottom panel of Table 1.

\subsection{Maximum disk fits}

\begin{figure*}
  \epsfxsize=\hsize \hfil\epsfbox{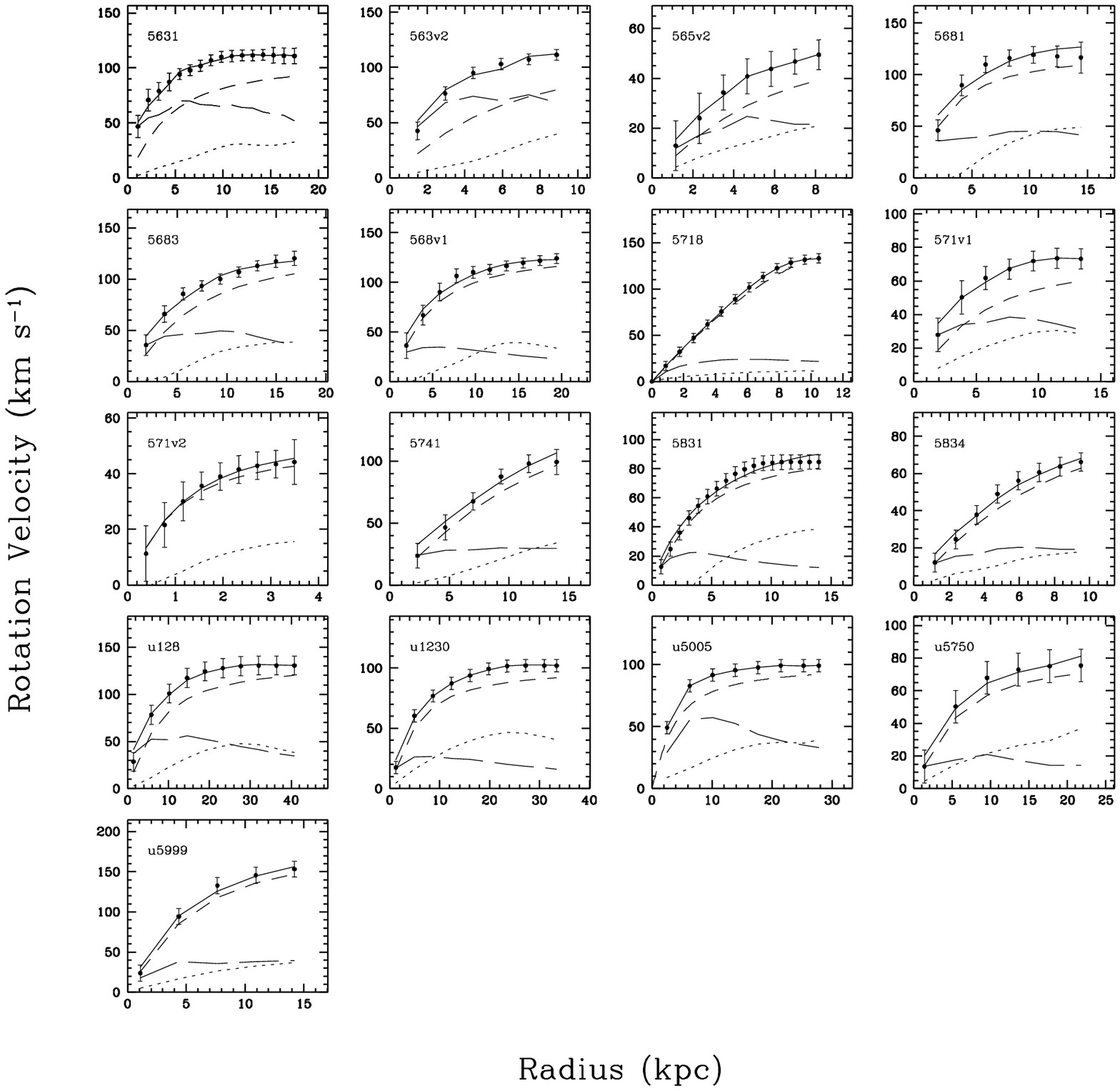}\hfil
\caption{Maximum disk rotation curve decompositions of the final sample
  of LSB galaxies. The dotted lines represent the rotation curves of
  the gas; the long dashed line those of the scaled stellar disk; the
  short dashed lines the rotation curves of the halo. The full line
  represents the total model rotation curve. Error bars are based on a
  combination of profile width in the position-velocity diagrams
  (BMH96) and the asymmetries between the rotation curves of both
  sides of the galaxies. Shown are the $R$ band decompositions, except
  for F563-V2 where we show the $B$ band data due to lack of $R$ data,
  and except for F571-V2 where the minimum disk solution is shown, as
  no optical photometry is available for that galaxy.}
\label{Maxdisk}
\end{figure*}
  
\begin{table*}
\begin{minipage}{160 mm}
\caption[]{Disk and halo parameters LSB galaxies}
\begin{tabular}{lrrrrrrrrrrrrrr}
  \hline &\multicolumn{6}{c}{Maximum disk
    $R$-band}&\multicolumn{6}{c}{Maximum disk $B$-band}\\ 
  \cline{2-7}\cline{9-14} \vspace{-8pt}\\ 
  Name&$\left({{M}\over{L}}\right)_{\star}$&$R_C$&$V_{\infty}$&$\rho_0$&
  $M_{\star}$&$M_{\rm
    d}$&&$\left({{M}\over{L}}\right)_{\star}$&$R_C$&$V_{\infty}$&$\rho_0$&
  $M_{\star}$&$M_{\rm d}$\\ (1)&(2)&(3)&(4)&(5)&(6)& (7)&& (8)& (9)&
  (10)&(11)&(12)&(13)\\ \hline
  F563-1 & 6.3 & 3.6 & 109.0 & 17.2 & 0.962 & 3.529 && 9.0 & 4.7 &
  106.8 & 9.6 & 1.374 & 3.063\\ 
  F563-V2 & * & * & * & * & * & * && 2.7 & 3.8 & 115.2 & 17.2 & 0.658
  & 1.456\\ F565-V2 & 3.1 & 4.6 & 61.8 & 3.4 & 0.103 & 0.310 && 3.6 &
  5.6 & 70.7 & 2.9 & 0.097 & 0.331\\ 
  F568-1 & 2.0 & 2.6 & 126.4 & 43.7 & 0.394 & 4.163 && 3.0 & 2.6 & 124.3 & 40.8 & 0.587 & 3.993\\ F568-3 & 1.8 & 5.4 & 136.5 & 11.8 & 0.429 & 4.190 && 2.6 & 6.6 & 147.4 & 9.1 & 0.532 & 4.305\\ F568-V1 & 2.0 & 3.8 & 135.9 & 23.6 & 0.207 & 5.929 && 2.6 & 4.2 & 137.2 & 19.5 & 0.332 & 5.824\\ F571-8 & 0.3\rlap{$^{a}$}& 7.9 & 249.9 & 18.2 & 0.097 & 9.934 && * & * & * & * & * & *\\ F571-V1 & 3.0 & 4.4 & 78.3 & 5.9 & 0.253 & 1.279 && 3.7 & 5.5 & 86.5 & 4.6 & 0.228 & 1.379\\ F574-1 & 0.9 & 10.2 & 172.1 & 5.3 & 0.216 & 3.681  &&         *  &       *   &       *  &       *  &        *   &       *\\ 
      F583-1   &      1.5  &     3.4  &     97.8  &     15.5  &      0.045  &    2.236  &&         *  &       *   &       *  &       *  &        *   &       *\\
      F583-4   &      0.9  &     6.0  &    104.2  &      5.6  &      0.065  &    0.962  &&         *  &       *   &       *  &       *  &        *   &       *\\
       U0128   &      3.0  &     6.4  &    136.3  &      8.3  &      1.043  &   14.32  &&       4.0  &     6.7   &   125.9  &     6.6  &    2.254   &  12.120\\
       U1230   &      0.7  &     4.5  &    102.7  &      9.5  &      0.187  &    6.912  &&      1.0  &     5.1   &   100.8  &     7.2  &    0.457   &   6.48\\
       U5005   &      4.5  &     3.2  &      91.8  &     15.0 &      0.675  &    4.467  &&         *  &       *   &       *  &       *  &        *   &       *\\
       U5750   &      0.7  &     4.5  &     83.4  &      6.4  &      0.086  &    2.526  &&         *  &       *   &       *  &       *  &        *   &       *\\
       U5999   &      3.0  &     4.6  &    190.7  &     32.0  &      0.444  &    7.966  &&         *  &       *   &       *  &       *  &        *   &       *\\ 
\hline
&&\multicolumn{4}{c}{Minimum disk}&&&\multicolumn{6}{c}{Bottema disk, $B$-band}\\
\cline{3-6}\cline{9-14}
\vspace{-8pt}\\  
Name &$M_{\rm g}$& $R_C$ & $V_{\infty} $ & $\rho_0$ & $M_{\rm d}$ &&& $\left({{M}\over{L}}\right)_{\star}$ & $R_C$ & $V_{\infty} $ & $\rho_0$ &
$M_{\rm d}$ & $M_{\star}$\\
(14)&(15)&(16)&(17)&(18)&(19)&&&(20)&(21)&(22)&(23)&(24)&(25)\\
\hline
      F563-1  &  0.385 &      1.5  &    118.6  &    111.2   &    5.042  &&&       1.7   &    1.8   &   113.4   &   75.4  &    4.507  &    0.203 \\
     F563-V2  &  0.321 &      1.9  &    127.6  &     80.9   &    2.479  &&&       1.3   &    2.8   &   118.9   &   34.4  &    1.858   &    0.365 \\
     F565-V2  &  0.084 &      2.8  &     59.1  &      8.4   &    0.399  &&&       1.3   &    3.4   &    59.1   &    5.6  &    0.355  &    0.038 \\
      F568-1  &  0.557 &      2.0  &    130.0  &     82.2   &    4.731  &&&       1.5   &    2.6   &   128.6   &   45.4  &    4.313  &    0.246 \\
      F568-3  &  0.394 &      3.4  &    131.2  &     28.2   &    4.742  &&&       1.6   &    5.2   &   135.7   &   12.5  &    4.219  &    0.308 \\
     F568-V1  &  0.344 &      3.1  &    134.3  &     35.0   &    6.153  &&&       1.4   &    3.7   &   135.5   &   25.5  &    5.972  &    0.128 \\
      F571-8  &  0.202 &      7.0  &     235.3  &     21.0   &    9.725  &&&       *   &      *   &       *   &      *  &        *  &        * \\
     F571-V1  &  0.164 &      2.5  &     79.9  &     18.6   &    1.640  &&&       1.4   &    3.3   &    78.2   &   10.7  &    1.448  &    0.103 \\
     F571-V2  &  0.016&      0.8  &     51.1  &     75.5   &    0.156  &&&         *   &      *   &       *   &      *  &        *  &        *\\
      F574-1  &  0.485 &      7.5  &    150.6  &      7.5   &    3.704  &&&       1.5   &   12.6   &   188.9   &    4.1  &    3.502  &    0.284 \\
      F583-1  &  0.243 &      2.8  &     96.5  &     21.8   &    2.322  &&&       1.5   &    3.4   &    97.9   &   15.3  &    2.235  &    0.036 \\
      F583-4  &  0.077 &      4.7  &     96.4  &      7.8   &    1.008  &&&       1.5   &    7.4   &   112.3   &    4.3  &    0.905  &    0.085 \\
       U0128  &  0.882 &      4.0  &    137.7  &     21.7   &   16.020  &&&       1.5   &    5.2   &   135.7   &   12.7  &   14.90  &    0.444 \\
       U1230  &  0.812 &      3.7  &    103.4  &     14.2   &    7.275  &&&       1.1   &    5.8   &   104.1   &    6.0  &    6.697  &    0.294 \\
       U5005  &  0.406 &      2.2  &    102.9  &     40.1   &    6.034  &&&       1.5   &    2.4   &    97.2   &   31.4  &    5.337  &    0.182 \\
       U5750  &  0.140 &      4.0  &     87.6  &      9.1   &    2.900  &&&       1.5   &    5.5   &    84.2   &    4.3  &    2.386  &    0.146 \\ 
       U5999  & 0.252 &      4.1  &    192.1  &     40.8   &    8.520  &&&       1.5   &    4.3   &   191.0   &   36.0  &    8.209  &     0.175 \\
\hline
\end{tabular}
Note: $(M/L)_{\star}$ is in units of $(M_{\odot}/L_{\odot})$ in the
respective bands; $R_C$ is in units of kpc, $V_{\infty}$ in units of
km s$^{-1}$, $\rho_0$ is in units of $10^{-3}\ M_{\odot} {\rm
  pc}^{-3}$; all masses are in units of $10^{10} M_{\odot}$. \\
{\it a:} $(M/L)_{\rm bulge} = 0$\\
\end{minipage}
\end{table*}   

\begin{table*}
\begin{minipage}{80 mm}
\caption[]{Maximum disk decompositions of HSB sample
  (Broeils 1992)}
\begin{tabular}{lrrrrrrr}
\hline
Name &$\left({{M}\over{L_B}}\right)_{\star}$& $R_C$ & $V_{\infty} $ &
$\rho_0$ & $M_{\rm d}$ & $M_{\star}$&$M_{\rm gas}$\\
\hline
      DDO154   &      1.2    &       2     &     59    &    15.3   &     0.39    &    0.01   &     0.04\\
      DDO168   &      1.1    &     2.7     &     98    &    24.2   &     0.21    &    0.02   &     0.03\\
      DDO170   &      2.7    &     2.3     &     75    &    19.2   &     0.85    &    0.03   &     0.06\\
         N55   &      0.7    &     7.9     &    146    &     6.4   &      1.5    &    0.29   &     0.13\\
        N247   &        4    &     7.3     &    136    &    6.54   &     1.33    &    0.95   &      0.1\\
        N300   &        2    &    6.32     &    132    &     8.3   &     1.74    &    0.42   &     0.16\\
        N801   &        4    &    74.3     &    302    &     0.3   &    19.14    &   23.83   &     2.86\\
       N1003   &      1.1    &     9.7     &    133    &     3.5   &     7.83    &    0.82   &     0.82\\
       N1560   &      3.5    &     6.8     &    133    &     7.2   &     0.94    &    0.12   &      0.1\\
       N2403   &      1.8    &     6.6     &    154    &    10.2   &     6.26    &    1.39   &     0.43\\
       N2841   &      5.1    &    21.7     &    308    &     3.7   &    116.1    &   23.97   &     2.09\\
       N2903   &      2.9    &     3.2     &    166    &      51   &    12.52    &    4.48   &     0.34\\
       N2998   &      2.1    &    24.8     &    242    &     1.8   &    26.95    &   13.79   &        3\\
       N3109   &      0.5    &     8.7     &    141    &     4.9   &     0.76    &    0.04   &     0.06\\
       N3198   &      3.5    &     7.6     &    156    &     7.8   &    11.25    &    3.11   &     0.69\\
       N5033   &      5.3    &     5.9     &    170    &    15.2   &    18.23    &    9.84   &     0.91\\
       N5533   &      6.8    &    34.6     &    255    &       1   &    52.89    &   30.05   &      3.2\\
       N5585   &      0.5    &     1.8     &     99    &    56.9   &     1.63    &    0.06   &     0.17\\
       N6503   &      1.7    &     2.5     &    115    &    38.6   &     5.74    &    0.83   &     0.22\\
       N6674   &        4    &   119.5     &    655    &     0.6   &    53.36    &   20.92   &      3.9\\
       N7331   &      5.8    &     103     &    982    &     1.7   &    32.42    &   12.95   &     1.58\\
       U2259   &      3.4    &     6.1     &    137    &     9.5   &     0.94    &    0.35   &     0.05\\
       U2885   &      2.1    &    44.9     &    382    &     1.3   &     1.87    &   37.37   &      4.8\\
\hline
\end{tabular}
Note: $(M/L_B)_{\star}$ is in units of $(M_{\odot}/L_{\odot ,B})$;
 $R_C$ is in units of kpc, $V_{\infty}$ in units of
km s$^{-1}$,  $\rho_0$ is in units of $10^{-3}\ M_{\odot} {\rm
  pc}^{-3}$; all masses are in units of $10^{10} M_{\odot}$. See
Broeils (1992) and references therein. 
\end{minipage}
\end{table*}

\begin{table*}
\begin{minipage}{120 mm}
\caption[]{Bottema disk and minimum disk decompositions HSB sample}
\begin{tabular}{lrrrrrrrrrrrrrrr}
\hline
&\multicolumn{6}{l}{\bf Bottema disk$^a$}&\multicolumn{4}{l}{\bf Minimum disk$^b$}\\
\hline
Name &$\left({{M}\over{L_B}}\right)_{\star}$& $R_C$ & $V_{\infty} $ &
$\rho_0$ & $M_{\rm d}$ & $M_{\star}$&$R_C$ & $V_{\infty} $ &
$\rho_0$ & $M_{\rm d}$\\
\hline
      DDO154   &    1.1    &   2.0     &  58.3    &   16.3   &    0.39   &   0.01&     1.7  &      56.3   &     21.1   &   0.39\\
      DDO168   &   1.0    &    2.0     &  79.3    &   29.7   &    0.23    &   0.002&       1.5  &      77.8   &     51.2   &   0.23\\
      DDO170   &        *    &       *     &      *    &       *   &        *    &       *&       1.6  &      70.6   &     35.4   &   0.84\\
         N55   &    1.3\rlap{$^d$}    &    6.4     &  140.4    &   9.0   &    1.53    &   0.26&       2.4  &     104.2   &     36.2   &    1.76\\
        N247   &    1.4    &    3.7     &   93.9    &   12.1   &    1.38    &   0.90&       1.4  &     107.3   &      114   &    2.11\\
        N300   &     1.6    &    6.9     &  134.2    &    7.1   &    1.70    &   0.46&       1.4  &     104.7   &      109   &    2.19\\
        N801   &    1.8    &    9.3     &  178.2    &   6.8   &    29.35    &   13.62&          --  &         --   &        --   &     --\\
       N1003   &    1.3    &    4.3     &  104.2    &   11.0   &    8.06    &   0.59&          --  &         --   &        --   &    --\\
       N1560   &    1.4    &    2.4     &  87.4    &   24.1   &    1.01    &   0.05&       1.8  &      84.1   &     40.6   &    0.96\\
       N2403   &    1.3    &    2.7     &  131.6    &   44.3   &     6.73    &    0.92&       0.8  &     134.3   &      470   &    7.95\\
       N2841   &     2.0    &    4.0     &  266.3    &   80.4   &    130.4    &   9.63&          --  &         --   &        --   &    --\\
       N2903   &    1.7    &    0.4     &  175.2    &    3933   &    15.03    &   1.97&          --  &         --   &        --   &    --\\
       N2998   &    1.5    &    1.5     &  177.6    &   262.8   &    34.16    &   6.58&          --  &         --   &        --   &   --\\
       N3109   &    1.5\rlap{$^d$}    &     8.7     &    141.0    &   4.9   &     0.76    &    0.04&       3.5  &      92.5   &     12.6   &   0.81\\
       N3198   &    1.4    &    1.4     &  138.7    &   189.6   &    12.88    &    1.48&          --  &         --   &        --   &   --\\
       N5033   &    1.5    &    0.04     &    181.0    &378800   &    24.28    &   3.79&          --  &         --   &        --   &    --\\
       N5533   &     2.0    &    0.2     &  190.9    &13930   &    69.18    &   13.76&          --  &         --   &        --   &    --\\
       N5585   &    1.4\rlap{$^d$}    &     1.8     &     99.0    &   56.0   &     1.63    &    0.06&       1.3  &      98.6   &      104   &    1.73\\
       N6503   &     1.6    &    0.6     &  110.8    &   610.4   &    6.24    &   0.34&          --  &         --   &        --   &    --\\
       N6674   &       --\rlap{$^c$}    &   23.6     &  247.1    &   2.0   &    59.93    &   14.34&          --  &         --   &        --   &    --\\
       N7331   &         --\rlap{$^c$}   &    9.3     &  233.4    &   11.8   &    38.55    &   6.82&          --  &         --   &        --   &    --\\
       U2259   &         --\rlap{$^c$}  &    1.2     &   81.2    &   89.1   &    1.07    &   0.22&       0.4  &      88.5   &      948   &    1.27\\
       U2885   &         --\rlap{$^c$} &   14.9     &  281.2    &   6.61   &    15.33    &   23.91&       1.0  &     287.9   &     1656   &    136.2\\
\hline
\end{tabular}
Note: $(M/L_B)_{\star}$ is in units of $(M_{\odot}/L_{\odot ,B})$ in the
respective bands; $R_C$ is in units of kpc, $V_{\infty}$ in units of
km s$^{-1}$,  $\rho_0$ is in units of $10^{-3}\ M_{\odot} {\rm
  pc}^{-3}$; all masses are in units of $10^{10} M_{\odot}$.\\
$a$:  The
Bottema disk decompositions were taken from Rhee (1996).\\
$b$: A dash indicates that a minimum disk fit was not possible.\\
$c$: $(M/L_B)_{\star}$ not determined because of unknown contribution
of bulge.\\
$d$: $(M/L_B)_{\star}$ is larger than maximum disk $(M/L_B)_{\star}$. 
\end{minipage}
\end{table*}

\begin{figure*}
  \epsfxsize=17cm \hfil\epsfbox{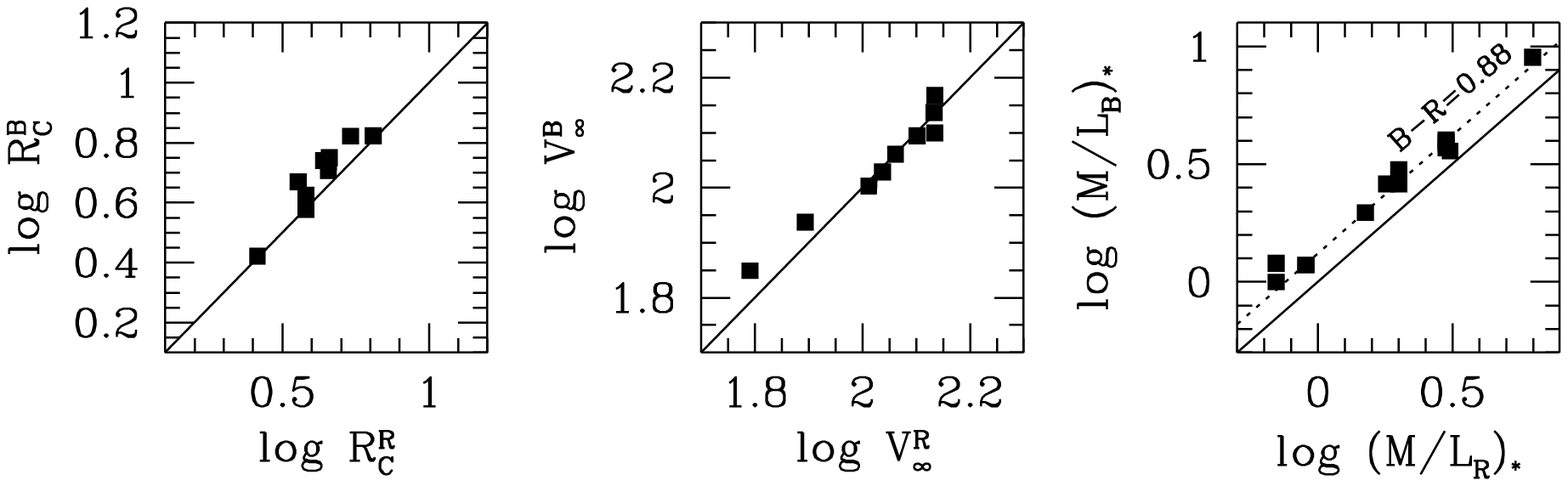}\hfil
\caption{Comparison of the core radii $R_C$ (left panel) and
  asymptotic velocities (middle panel) $V_{\infty}$ of the halos, and
  the maximum disk mass-to-light ratio $(M/L)_{\star}$ of the stellar
  disk (right panel) resulting from separate $B$ and $R$ band
  decomposition of the rotation curves of the subset of LSB
  galaxies for which both $R$ and $B$ band photometry were available.
  The $B$ and $R$ band decompositions are consistent with each other.
  The full drawn lines are lines of equality. The dotted line in the
  right panel is the relation between $(M/L_B)_{\star}$ and
  $(M/L_R)_{\star}$ for $B-R = 0.88$.}
\label{compare}
\end{figure*}

The maximum disk hypothesis attributes as much of the observed
rotation velocity to the stellar disk as possible. It therefore always
yields an upper limit to the $(M/L)_{\star}$ in a galaxy. For HSB
galaxies the maximum disk hypothesis can explain almost all of the
observed rotation velocity in the inner parts with just the stellar
disk. For LSB galaxies the situation is quite different.  Figure
\ref{Maxdisk} shows the maximum disk decompositions of their rotation
curves.  The resulting disk and halo parameters are summarized in
Table 2, together with the masses of the various components, as
measured within the outermost radius $R_{\rm max}$ of the rotation
curve (Table 1).  For those galaxies where both $R$ and $B$ photometry
were available we have done separate decompositions for both bands,
which are compared in Fig.  \ref{compare}.  The core radii and
asymptotic velocities correspond very well to each other in both sets,
as do the maximum disk $(M/L)_{\star}$ ratios when the $B-R$ colors
of the galaxies are taken into account.  To compare those LSB galaxies
for which only $R$-band data is available with $B$-band literature
data, we will convert the measured maximum disk $(M/L_R)_{\star}$
ratios to $B$-band values assuming an average $B-R$ color of 0.88 (de
Blok et al.\ 1995, cf.\ Fig.\ \ref{compare}). Our results do not
depend on this conversion as the halo is the most dominant component
and its parameters are therefore not very sensitive to the precise
value of $(M/L)_{\star}$.  The maximum disk decompositions for the HSB
samples have been taken from Broeils (1992) and are for convenience
summarized in Table~4.

\begin{figure}
  \epsfxsize=8.5cm \hfil\epsfbox{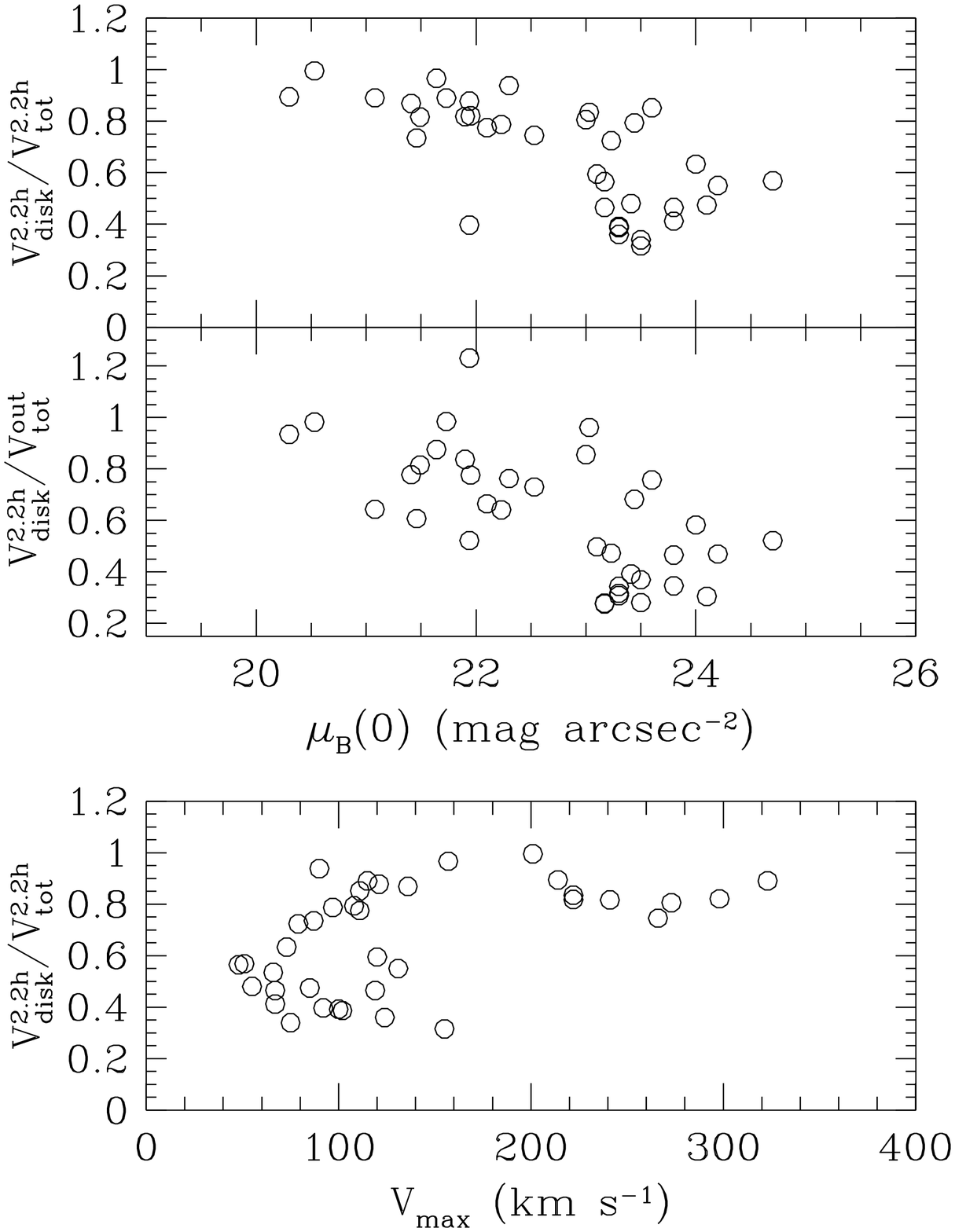}\hfil
\caption{Top panel: the ratio of the peak 
  rotation velocity of the maximum disk and the observed total
  rotation velocity at 2.2$h$. This ratio is approximately constant at
  $\sim 0.85$ for galaxies with $\mu_0(B) < 23$ mag arcsec$^{-2}$ and
  drops steeply to values of $\sim 0.5$ for the LSB galaxies. Even at
  2.2$h$ the importance of the disk decreases toward lower surface
  brightnesses, despite higher values of $(M/L)_{\star}$. Middle
  panel: the ratio of the peak rotation velocity of the maximum disk
  and the observed total rotation velocity at the outermost measured
  point. This ratio also changes systematically with surface
  brightness, from $\sim 1$ for the highest surface brightness
  galaxies, where the disk determines the maximum rotation velocity,
  to $\sim 0.3$ for LSB galaxies, where the halo completely determines
  the dynamics of these galaxies.  Bottom panel: the ratio of maximum
  disk peak velocity and observed velocity at 2.2$h$. At $V_{\rm max} <
  150$ km s$^{-1}$ a range of ratios is observed, from $\sim 0.85$ for
  HSB galaxies to $\sim 0.4$ for LSB galaxies. The lack of points with
  low ratios at $V_{\rm max} > 150$ km s$^{-1}$ might be caused by
  selection effects: LSB galaxies with such large rotation velocities
  have not yet been mapped in H{\sc i}.}
\label{ratio}
\end{figure}

\subsection{Minimum disk}

Many of the rotation curves of LSB galaxies are best fitted by a
minimum disk, that is, under the assumption that $(M/L)_{\star}=0$.
We have calculated the halo parameters under this assumption (Table
3).  We will compare them with the maximum disk parameters in Section
5.4, but already note that there is little change in the structural
parameters of the halos, due to the DM dominance.  Although minimum
disk is not a realistic assumption for HSB galaxies, we have for
comparison also made minimum disk fits to the rotation curves of the
HSB sample (Table 5).  Not all galaxies could be fitted: especially
the earlier types (with bulges) have rotation curves that rise too
steeply to be fitted by a simple isothermal model with a finite core
radius, thus clearly showing the need for a compact luminous
component.

\subsection{Bottema disk}

Bottema (1995) has measured the stellar velocity dispersions of a
sample of HSB spiral galaxies and found that the dispersion is related
to the maximum rotation velocity.  For a Freeman disk with a color
$B-V = 0.7$ he derived $(M/L_B)_{\star}\simeq 1.8$.  The stellar disks
of the galaxies in his sample supply approximately 63\% of the total
measured velocity at 2.2$h$ (where the rotation curve of an
exponential disk peaks).  These results were derived under the
explicit assumption of Freeman's Law (which holds for most of
Bottema's galaxies).  It is obvious that Bottema's original
prescription cannot hold in LSB galaxies: 63\% of the total rotation
velocity is in many cases already more than the maximum disk solutions
allow.  In Appendix A we give a more general expression for
$(M/L_B)_{\star}$ of a galactic exponential disk derived by Bottema
(1997).  The resulting $(M/L_B)_{\star}$ only depends on color, as
the effects of luminosity and surface brightness cancel in this model.
In the absence of data on stellar velocity dispersions of low or
intermediate surface brightness galaxies, this is the best we can
currently do taking into account the actual gravitational potential of
the disk.  In the following we will refer to this prescription as the
``generalized Bottema-disk.'' Of course, when applying this
generalized scheme to HSB galaxies, the original 63\% criterion is
retrieved.

The $(M/L)_{\star}$ values given in column (20) of Table 3 show that
the range of values implied by the Bottema disk is very small compared
to the maximum disk values.  This is due to the relatively small range
in $B-V$ colors shown by the galaxies.  We have used Bottema's
original ``63\% recipe'' (as presented in Rhee 1996) for galaxies with
$\mu_0^B < 23 {\rm\ mag\ arcsec}^{-2}$, while the generalized Bottema
disk was used for galaxies with $\mu_0^B$ fainter than this value,
where 63\% of the total rotation velocity is already more than maximum
disk allows.  In practice the generalized disk was thus used for the
LSB galaxies, NGC 247 and the three DDO galaxies.  For NGC 55, NGC
3109 and NGC 5585 the Bottema recipes resulted in $(M/L_B)_{\star}$
values larger than those implied by maximum disk and we have assumed
the maximum disk values.  In at least one case (NGC 3109) it can be
shown that the discrepancy arises because of a large difference between
the optical (Fabry-Perot) and \HI\ rotation curves. If the Fabry-Perot
curve is used the  maximum disk $(M/L_B)_{\star}$ value becomes larger
than the Bottema value. 

\section{Results of decompositions}

\subsection{The influence of resolution on disk-halo decompositions}

As we have established in Section 3 that beam-smearing is not
determining the shapes of the rotation curves, we should not expect
the results of disk-halo decompositions of the LSB galaxies to be
influenced significantly by resolution effects.  We test this by again
looking at the case of NGC 1560 (Figure \ref{smoreg1560}).

Even the high-resolution decompositions of this galaxy are not
entirely unambiguous, showing that even in a maximum disk
decomposition it is difficult to entirely constrain the
$(M/L)_{\star}$ value.  The quoted maximum disk $(M/L_B)_{\star}$
values range from 3.5 (Broeils 1992, Chapter 10) and 4.1 (Begeman,
Broeils \& Sanders 1991) to 4.9 (Broeils 1992, Chapter 5). The latter
case is however an attempt to fit the data without a dark halo, and we
will discard that value.  To be consistent with Tables 1 and 4 we
adopt $(M/L_B)_{\star} = 3.5$.  This maximum disk decomposition is
shown in the left panel of Fig.  \ref{smoreg1560}.  Corresponding halo
parameters, taken from Table 3 are repeated in Table 6.

As described in Sect.\ 3.3, the rotation curves of NGC 1560 were
smoothed from a resolution of 38 beams to 3 beams. We have attempted a
maximum disk decomposition on this low resolution data, and find that,
although again there is some freedom in $(M/L)_{\star}$, the
best-fitting value for the low-resolution data is $\simeq 4$.  We can
exclude $(M/L_B)_{\star}$ = 3 or 5 as possible solutions for this
particular case, as the corresponding low-resolution models
systematically over- or under-estimate the rotation velocities at
certain radii.  The low-resolution halo parameters are given
Table 6, and are comparable with the high-resolution values.

\begin{table}
\begin{minipage}{60 mm}
\caption[]{Halo parameters of NGC 1560}
\begin{tabular}{lll}
  \hline  
 & high res. & low res. \\ 
\hline 
$(M/L_B)_{\star}$ & 3.5  & 4   \\ 
$V_{\infty}$    & 133  & 121   \\ 
$R_C$           & 6.8  & 6.2    \\
$\rho_0$        & 7.2  & 5.9   \\
\hline
\end{tabular}
\end{minipage}
\end{table}

The value $(M/L_B)_{\star} = 4$ for the low-resolution case was
determined by simply scaling the stellar rotation curve to match the
innermost point. This procedure is thus for the low resolution case
equivalent to a formal maximum disk fit in the high-resolution case.
Studying high-resolution curves in detail shows that this scaling to
match the innermost point is usually what happens in practice when
doing a formal maximum disk fit.
        
\subsection{Maximum disk mass-to-light ratios}

As is clear from Fig. \ref{Maxdisk}, the rotation curves of the
stellar disks in the LSB galaxies (even in the maximum disk case) do
not even approach the maximum rotation velocities of the observed
rotation curves. This is in sharp contrast with HSB galaxies where the
rotation curve of the stellar disk is usually able to explain most of
the rotation velocity in the inner parts.

This is also illustrated in the top panel of Fig.  \ref{ratio}. This
shows the ratio of the peak velocity of the maximum disk at 2.2$h$ and
the total observed rotation velocity at that radius. At central
surface brightnesses brighter than 23 $B$-mag arcsec$^{-2}$, the
maximum disk can account for 85\% $\pm$ 10\% of the total rotation
velocity, but this rapidly drops to only 40\% or less for LSB
galaxies.

This increasing inability of the maximum disk to explain the observed
rotation velocities is illustrated in the middle panel of Fig.
\ref{ratio}, which shows the ratio of the peak velocity of the maximum
stellar disk and the maximum observed rotation velocity.  This ratio
systematically changes from $\sim 0.9$ at high surface brightnesses
(i.e.  the disk defines the maximum rotation velocity and the halo
``conspires'' to keep the rotation curve flat) to values of $\sim 0.3$
for the LSB galaxies (i.e.  the halo completely determines the
dynamics of the galaxies). The bottom panel of Fig. \ref{ratio} again
shows the ratio of of the peak velocity of the maximum disk at 2.2$h$ and
the total observed rotation velocity at that radius, but as a function
of $V_{\rm max}$. 

\begin{figure*}
  \epsfxsize=\hsize \hfil\epsfbox{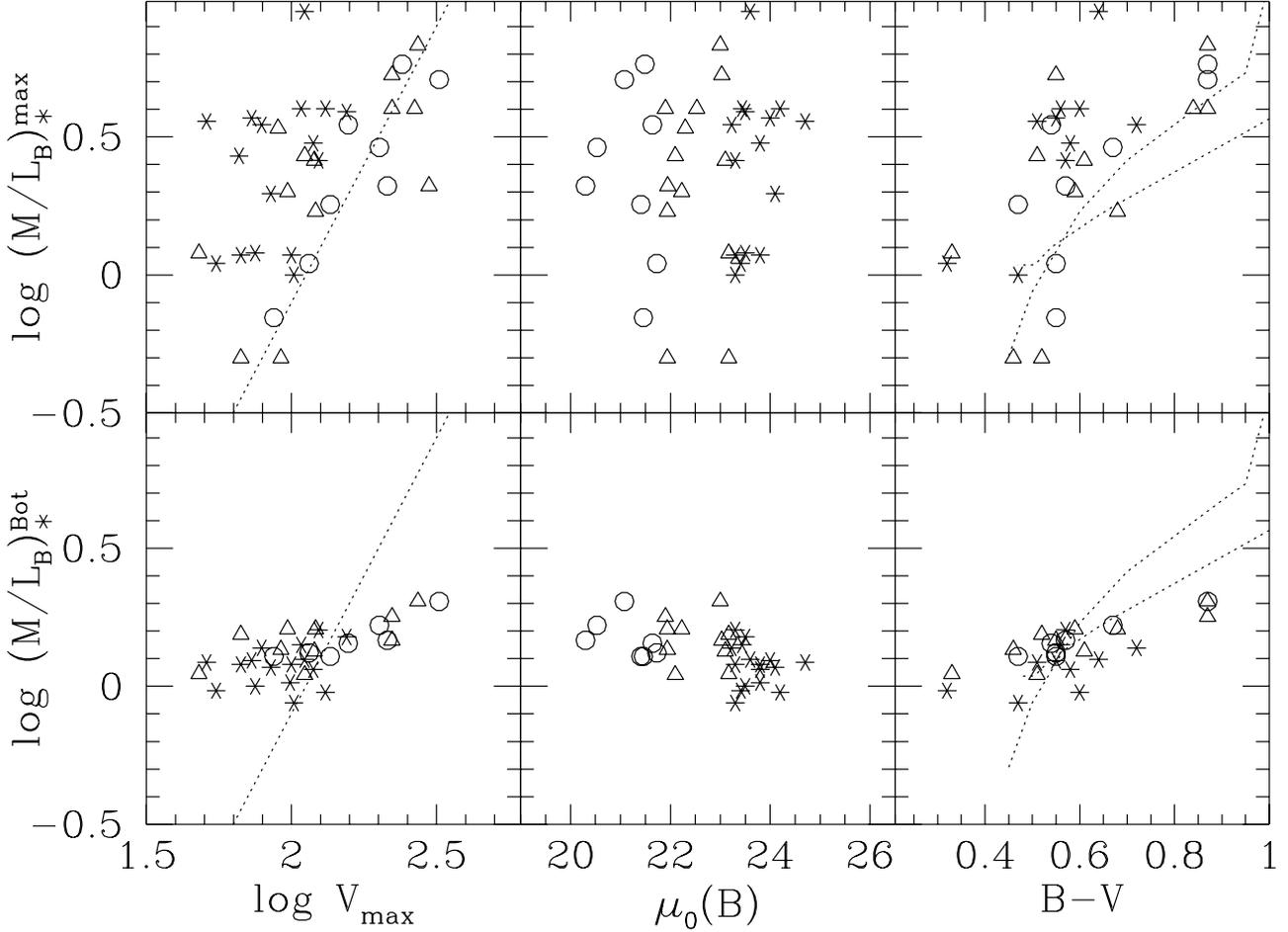}\hfil
 \caption{{\bf Top row:} The maximum disk $B$-band stellar 
   mass-to-light ratio plotted versus maximum rotation velocity (left
   panel), central surface brightness (middle panel), and average
   $B-V$ color (right panel).  The open circles denote galaxies with
   $\mu_0(B) < 21.9$ mag arcsec$^{-2}$, the triangles represent
   $21.9 \le \mu_0(B) < 23.2$ mag arcsec$^{-2}$ galaxies and the
   asterisks galaxies with $\mu_0(B) \ge 23.2$ mag arcsec$^{-2}$.  In
   the left panel the dotted line shows the relation $(M/L_B)_{\star}
   \propto L^{1/2}$. The relation has been arbitrarily shifted to
   follow the filled HSB data points. In the right panel the steep
   upper dotted curve shows the colour--mass-to-light ratio track from
   Larson \& Tinsley (1978) while the shallow lower dotted curve shows
   the track as derived by Jablonka \& Arimoto (1992). Both curves
   have been shifted to go through $\log (M/L_B)_{\star} = 0.55$ at
   $B-V = 0.55$ (see bottom-right panel).  {\bf Bottom row:} Idem, but
   now showing the $B$-band stellar mass-to-light ratio as derived
   assuming a Bottema disk. }
\label{MLBmax} 
\end{figure*}

The maximum disk $(M/L)_{\star}$ values are shown in Fig.~\ref{MLBmax}
as a function of $\mu_0^B$, $V_{\rm max}$ and $B-V$.  The relation
between $(M/L_B)_{\star}$ and $V_{\rm max}$ (top-left) shows a clear
segregation between galaxies of different surface brightnesses: at
each value of $V_{\rm max}$ the HSB galaxies have the lowest value of
$(M/L_B)_{\star}$ while the LSB galaxies always show high
$(M/L_B)_{\star}$ values.  Also shown is the relation $(M/L_B)_{\star}
\propto L^{1/2}$ which was previously derived for HSB galaxies by
e.g.  Ashman, Salucci \& Persic (1993) and Broeils (1992).  Clearly
the conclusion that late-type (i.e. lower surface brightness) galaxies
have smaller maximum disk stellar mass-to-light ratios than early-type
(i.e. higher surface brightness) galaxies was, in the light of this
new data, a selection effect.  The maximum disk $(M/L_B)_{\star}$ is
not a simple function of $L$ (or $V_{\rm max}$) alone, {but also
depends on surface brightness}.

The absence of a relation between $(M/L_B)_{\star}$ and $\mu_0$
(top-center) is a result of comparing galaxies with different global
properties: at each $V_{\rm max}$ there is a large range in $\mu_0$
(and consequently a large range in $(M/L_B)_{\star}$). Comparing
galaxies at fixed $\mu_0$ means comparing galaxies of different masses
and luminosities.

The surface brightness segregation effect at fixed $V_{\rm max}$
cannot be an artificial effect caused by the lower resolution of the
LSB observations.  We have shown in Section 3 that beam smearing
effects are potentially most severe in HSB galaxies.  If beam-smearing
were to have affected the LSB curves this would imply even higher
values for $(M/L)_{\star}$, thus making the difference between HSB and
LSB galaxies seen in Fig.\ \ref{MLBmax} even more dramatic (see also
discussion in de Blok \& McGaugh 1996).

\subsubsection{The maximum exponential disk}
The segregation in $(M/L)_{\star}$ can be explained from the
properties of the exponential disk and the systematics of maximum disk
fitting.  As LSB galaxies obey the same Tully-Fisher relation as HSB
galaxies (Zwaan et al.\ 1995), this requires LSB galaxies to have
larger scale lengths at fixed $V_{\rm max}$, in order to still reach
the required value of $L$.  The maximum rotation velocity of an
exponential disk is given by $v \propto \sqrt{\sigma_0 h}$, where
$\sigma_0$ is the central mass surface density. In practice the value
of $\sigma_0$ is not known, and the distribution of the light is used
to compute the rotation curve of the disk. This rotation curve of the
light has a maximum rotation velocity of
\begin{equation}v_{1} \propto \sqrt{\Sigma_0 h},
\end{equation}  
where $\Sigma_0$ is the central surface brightness in linear units.
(In the following we will denote the rotation velocity of the disk with
{\it v}, while the observed total rotation velocity will be indicated with
{\it V}.)
In the rotation curve fitting process the mass-to-light ratio
$(M/L)_{\star}$ is introduced to convert surface brightness to mass
surface density and compute the disk rotation curve. Under the maximum
disk hypothesis $(M/L)_{\star}$  is scaled to its maximum possible
value.  The peak rotation velocity of the maximum disk is then given
by
\begin{equation}
  v_{\rm m} \propto\sqrt{(M/L)_{\star}\Sigma_0 h}.
\end{equation}
From Eqs. (5) and (6) it is clear that
\begin{equation}
  \left( {{M}\over{L}} \right)_{\star} = \left( {{v_{\rm
          m}}\over{v_1}} \right)^2.
\end{equation}

We will now compare a HSB and a LSB galaxy at identical positions on
the Tully-Fisher relation. We find for the peak velocity of the
rotation curve of the light in the HSB, $v_1^{\rm H}$, that $(v_1^{\rm
  H})^2 \propto ({\Sigma_0^{\rm H} h^{\rm H}})$, and for the peak
velocity in the LSB, $v_1^{\rm L}$, that $(v_1^{\rm L})^2 \propto
({\Sigma_0^{\rm L}}h^{\rm L})$, where $\Sigma_0^{\rm L,H}$ and $h^{\rm
  L,H}$ are the central surface brightnesses and scale lengths of the
LSB and the HSB, respectively.  As their luminosities $L \propto
\Sigma_0 h^2$ are identical we find $h\propto \Sigma_0^{-\frac{1}{2}}$
and we can write $v_1^{\rm L}$ as
\begin{equation} (v_1^{\rm L})^2 \propto {\Sigma_0^{\rm L}h^{\rm L}}
  \propto {{{\Sigma_0^{\rm L}}\over{\Sigma_0^{\rm H}}} \Sigma_0^{\rm H} \left({
      {{\Sigma_0^{\rm H}}\over{\Sigma_0^{\rm L}}}}\right)^{\frac{1}{2}} h^{\rm H}} \propto
  \left({{{\Sigma_0^{\rm L}}\over{\Sigma_0^{\rm H}}}}\right)^{\frac{1}{2}} (v_1^{\rm H})^2.
\end{equation}
Using Eqs. (7) and (8) we express $(M/L)_{\star}^{\rm L}$ in term of
$(M/L)_{\star}^{\rm H}$.

\begin{eqnarray}\nonumber
  \left( {{M}\over{L}} \right)_{\star}^{\rm L} & 
= & \left( {{v_{\rm m}^{\rm L}}\over{v_1^{\rm L}}} \right)^2
=\left({{\Sigma_0^{\rm H}}\over{\Sigma_0^{\rm L}}}\right)^{\frac{1}{2}}
  \left( {{v_{\rm m}^{\rm L}}\over{v_1^{\rm H}}} \right)^2\\ 
&=&\nonumber\left({{\Sigma_0^{\rm H}}\over{\Sigma_0^{\rm L}}}\right)^{\frac{1}{2}}\left(
    {{v_{\rm m}^{\rm H}}\over{v_1^{\rm H}}} \right)^2\left( {{v_{\rm
          m}^{\rm L}}\over{v_{\rm m}^{\rm H}}} \right)^2 \\
& =&\left({{\Sigma_0^{\rm H}}\over{\Sigma_0^{\rm L}}}\right)^{\frac{1}{2}}
  \left( {{v_{\rm m}^{\rm L}}\over{v_{\rm m}^{\rm H}}} \right)^2\left(
    {{M}\over{L}} \right)_{\star}^{\rm H}.
\end{eqnarray}

For galaxies at identical positions on the Tully-Fisher relation, the
ratio of their maximum disk $(M/L)_{\star}$ values thus depends on the
ratio of the surface brightnesses and the ratio of their peak rotation
velocities.  This latter ratio depends on the shape of the rotation
curves or, more specifically, the change of shape with surface
brightness at fixed $V_{\rm max}$.  We have already seen that the
rotation curves of LSB galaxies rise more slowly than those of HSB
galaxies, so we expect the peak velocities of the maximum disk of the
LSB galaxies to be smaller than those of the HSB galaxies.

In order to get $(M/L)_{\star}^{\rm L} > (M/L)_{\star}^{\rm H}$,
Eq. (9) requires that
\begin{equation} \left({{v_{\rm m}^{\rm L}}\over{{v_{\rm m}^{\rm
          H}}}}\right)^4 \ge {{\Sigma_0^{\rm L}}\over{\Sigma_0^{\rm
          H}}}.
\end{equation} 
It is easy to show that if both LSB and HSB galaxies had flat rotation
curves, we find that $(M/L)_{\star}^{\rm L} > (M/L)_{\star}^{\rm H}$
always. In practice, at each $V_{\rm max}$ a large range of
$v(2.2h)/V(2.2h)$ is found (Fig.\ 7).  For the current sample, where a typical
HSB galaxy has $\mu_0 \simeq 21.5$, while a typical LSB galaxy has
$\mu_0 \simeq 24$, we find that condition (10) requires $v_{\rm
  m}^{\rm L}/v_{\rm m}^{\rm H} \ga 0.5$ on average.  We find the
largest range of $v(2.2h)/V(2.2h)$ at $V_{\rm max} \simeq 100$ km
s$^{-1}$ in the bottom panel of Fig. 7. Taking the ratio of the
average value for the LSB galaxies $v(2.2h)/V(2.2h) \simeq 0.45$ and
that of the HSB galaxies $v(2.2h)/V(2.2h) \simeq 0.85$, yields a value
of $v_{\rm m}^{\rm L}/v_{\rm m}^{\rm H} \simeq 0.5$. So although there
will be individual cases where condition (10) will not hold (as is
shown by the overlap of HSB and LSB points in the top-left panel of
Fig. 8), {\it on average} condition (10) is always fulfilled, and we
find that the intrinsic properties of the exponential disk combined
with the maximum disk recipe always yield larger values of
$(M/L)_{\star}$ for LSB galaxies than for HSB galaxies. No assumptions
have been made on the stellar populations, ages, or evolution of the
galaxies.  The larger {\it fitted} maximum disk stellar mass-to-light
ratios for LSB galaxies at fixed $V_{\rm max}$, therefore do not need
to reflect the true evolutionary stellar mass-to-light ratio of the
disk.

\subsubsection{Population effects?}

But is it in fact at all possible to explain the systematically higher
maximum disk $(M/L)_{\star}$ ratios of the LSB galaxies at fixed
$V_{\rm max}$ in a consistent way by invoking systematic changes in
the stellar populations?
The systematically larger ratios imply an increasingly
more evolved population with decreasing surface brightness: population
synthesis models (Larson \& Tinsley 1978) show that $(M/L_B)_{\star}$
of an evolving stellar population is in general an increasing function
of time, and hence of increasing $B-V$ color.

Two examples are shown in in the right panels of Fig.
\ref{MLBmax}, where the upper dotted line represents $(M/L_B)_{\star}$ as a
function of the $B-V$ color of a stellar population
with a declining star formation rate 
(Larson \& Tinsley 1978).  
The lower dotted line is the track as derived by Jablonka \& Arimoto
(1992) who have used a model which also takes into
account chemical enrichment.

To a certain degree the maximum disk results agree with the trend of
increasing $(M/L_B)_{\star}$ with redder $B-V$: the lower envelope of
the distribution of the points increases with redder color.
Uncertainties in the low-mass end of the IMF make it very hard to
derive absolute $(M/L_B)_{\star}$ values from any population synthesis
model. This uncertainty may therefore shift the model track up or
down, while e.g.  metallicity effects may make the slope of the track
more shallow (as the Jablonka \& Arimoto track shows).  However, the
observed range at each color is too large to be explained by models
with a constant IMF and where the star formation history is
well-behaved and a smoothly varying function of Hubble type.

The high maximum disk $(M/L)_{\star}$ are furthermore in direct
conflict with observational evidence on colors (McGaugh \& Bothun 1994, de Blok et al. 1995),
metallicities (McGaugh 1994) and gas-fractions (McGaugh \& de Blok
1997).  As the stellar populations of LSB galaxies are in an earlier
evolutionary state than HSB galaxies (van den Hoek et al.  1997) one
would expect them to have {\it lower} values of $(M/L)_{\star}$.  The
values of $(M/L_B)_{\star}$ as derived under the maximum disk
hypothesis are therefore {\it not} representative for the stellar
population in LSB galaxies.

\subsubsection{Disk dark matter?}

Trying to explain the maximum disk results in a physical way, while
having a stellar population with a lower $(M/L)_{\star}$ than implied
by the maximum disk values, requires an additional dark component in
the disk which traces the stellar population. This component has to
become increasingly important towards lower central surface
brightnesses.  This might be an invisible baryonic component like e.g.
cold molecular gas.  There has already been much speculation about
this in the literature (e.g.  Pfenniger, Combes \& Martinet 1994).
LSB galaxies would be the ideal test cases for such theories as the
major part of their disks should then be made up of this baryonic DM.
Again, this component should be of increasing importance towards lower
surface brightnesses.

This creates the unattractive picture of two dominant but distinct and
differently distributed dark components: one that has to provide the
extra mass in the disk to make it satisfy the maximum disk boundary
condition, and one that has to provide the surplus velocity in the
outer parts to produce the observed rotation curves.  These two
dominant components then have to ``conspire'' to produce the observed
 rotation curve. Either way, DM remains of extreme
importance in LSB galaxies.

\subsubsection{Non-maximum disk}

Another possibility is that disks are simply non-maximal. Part of the
halo DM is wrongly attributed to the disk simply because the disk does
not dominate the dynamics in the inner parts.  This could in principle
affect the derived values for the structural parameters of the halo,
but this is only of minor importance in LSB galaxies. They are DM
dominated, irrespective of the value of $(M/L)_{\star}$, so that
changing the mass of the disk does not really matter.  It is, however,
of crucial importance in HSB galaxies where the disk is dominant, and
the halo parameters therefore sensitive to the precise value of
$(M/L)_{\star}$.

The maximum disk hypothesis, in combination with the extendedness of
LSB galaxies thus produces a balancing act where LSB galaxies have
to have large $(M/L)_{\star}$ ratios, or extra dark components, which
are contrary to what we can derive from other evidence (stellar
populations, metallicity, gas fraction). Maximum disk is therefore not
the preferred way of making mass models in LSB galaxies.

\subsection{Maximum disk halo parameters}

\begin{figure*}
\epsfxsize=17cm
\hfil\epsfbox{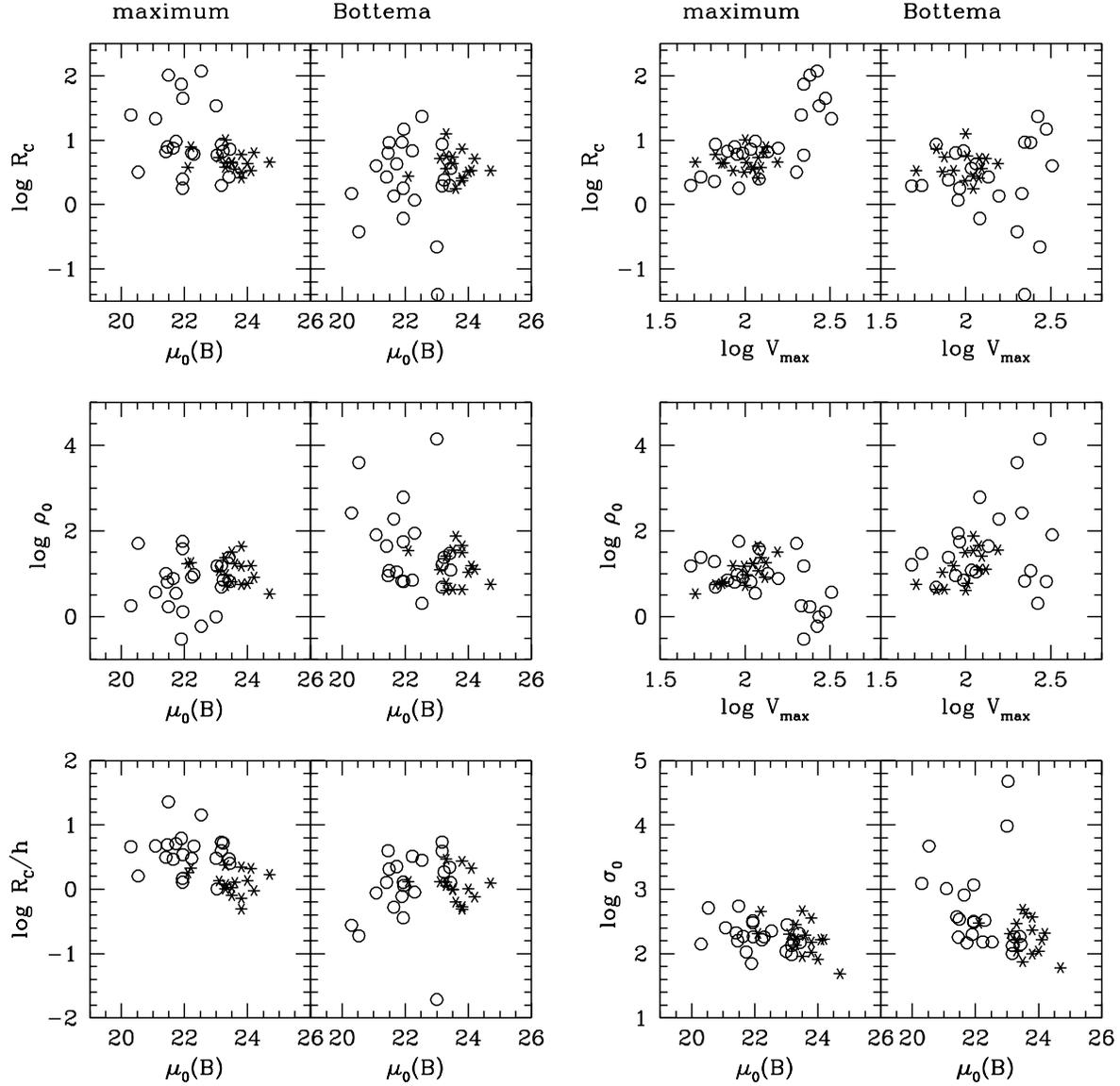}\hfil
\caption{Isothermal halo fitting parameters for maximum disk fits
  (left panels) and Bottema disk fits (right panels).  The open circles
  represent the HSB sample, the asterisks the LSB sample.  $\rho_0$ is
  expressed in units of $10^{-3} $M$_{\odot}$ pc$^{-3}$; $R_C$ in kpc;
  $\sigma_0$ in $10^{-3}$ M$_{\odot}$ pc$^{-2}$; $V_{\rm max}$ in km
  s$^{-1}$; and $\mu_0(B)$ in mag arcsec$^{-2}$.}
\label{Vmaxtot}
\end{figure*}

For the LSB galaxies the asymptotic velocity $V_{\infty}$ is well
correlated with the observed maximum rotation velocity $V_{\rm max}$,
showing that although many of the LSB curves are still rising in the
outermost point, they are close to their true maximum velocity.  We
will use this observational parameter instead of the fitted
$V_{\infty}$.  Figure \ref{Vmaxtot} summarizes the halo parameters for
the HSB and LSB galaxies under the assumption of maximum disk.  The
core radius $R_C$ is rather constant, both as a function of $V_{\rm
  max}$ and $\mu_0$.  The group of points at large $R_C$ belong to HSB
galaxies that are very much dominated by luminous matter, and are
maximum disk in the true sense of the word: their inner rotation curve
can be explained completely by that of the disk. To avoid having a
hollow halo in these galaxies, the maximum disk $(M/L)_{\star}$ is
usually slightly, but somewhat arbitrarily, lowered. The halo
parameters are extremely sensitive to this.  In most cases these
galaxies have a bulge, that also has to be accommodated in the fitting
procedure, with its own uncertainties in its $(M/L)_{\star}$.  The HSB
points at large $R_C$ are thus uncertain.

The central halo densities $\rho_0$ as a function of $\mu_0$ tend to
be somewhat higher than those of HSB galaxies, but again, the
uncertainty in the decompositions of the HSB galaxies makes it
difficult to say anything more definite.
 
The ratio between core radius and optical scale length changes with
surface brightness from $R_C \simeq 3 h$ for the HSB galaxies to $R_C
\simeq h$ for the LSB galaxies.  In the maximum disk picture the
optical disks of LSB galaxies are massive and 
extend further out into the halo.

\subsection{Minimum disk}

\begin{figure}
  \epsfxsize=8.5cm \hfil\epsfbox{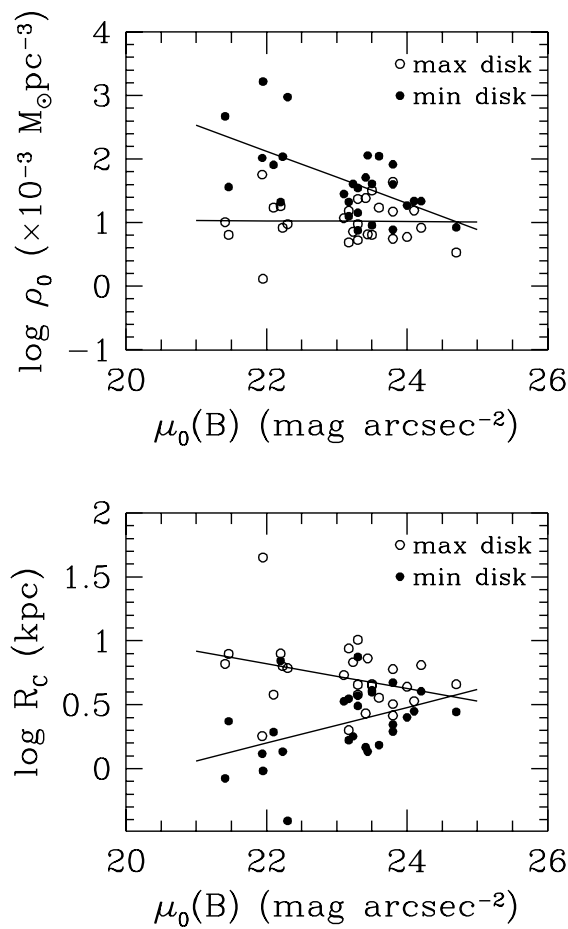}\hfil
\caption{The central densities (top) and core radii (bottom)
of the halos HSB and LSB galaxies
  under the maximum disk assumption (open circles) and minimum
  disk assumption (filled circles). The core radii and densities of
  the HSB galaxies are more affected by the respective assumptions
  than those of the LSB galaxies. Note the different slopes of the
  trends under the different hypotheses. The drawn lines are
  least-squares fits.}
\label{Minmaxrc}
\end{figure}

We can illustrate that the halo parameters derived for the LSB
galaxies are robust values, by comparing the values derived for
maximum disk and minimum disk. This is done in Fig. \ref{Minmaxrc},
where the core radii and central densities as derived using these two
extreme hypotheses are compared.  The difference in maximum and
minimum disk halo parameters is clearly a strong function of surface
brightness.  The largest difference is observed for the HSB galaxies,
where the central density $\rho_0$ changes by more than an order of
magnitude. A similar conclusions applies to the core radius.
This is in sharp
contrast with the core radii derived for the LSB galaxies. The minimum
disk values differ from the maximum disk values by less than a factor
of two.  The slope of the trend with surface brightness changes from
negative to positive when going from maximum disk to minimum disk.
This is entirely caused by the large change in parameters of the HSB
galaxies.

The DM dominance makes the parameters of the LSB galaxy halos
insensitive to the precise $(M/L)_{\star}$ value. The values we derive
for the halo structural parameters, but also for the dark mass in
these halos, are therefore likely to be very close to their true
values.

\subsection{Bottema disk mass-to-light ratios}

The most important property which distinguishes the Bottema disk from
the maximum disk is its small range of $(M/L_B)_{\star}$.  This is
immediately apparent in Fig.  \ref{MLBmax}.  The Bottema disk
typically implies values of $(M/L_B)_{\star}$ between 1 and 2.  In
general the reddest galaxies have the highest mass-to-light ratios.

The striking systematic offset in $(M/L)_{\star}$ at fixed $V_{\rm
  max}$ between HSB and LSB galaxies for the maximum disk hypothesis
has disappeared (compare with top left panel in Fig.~\ref{MLBmax}{}).
At fixed $V_{\rm max}$ galaxies now have approximately identical
$(M/L)_{\star}$ ratios.  The majority of the data points has a much
smaller spread and is in much closer agreement with 
stellar population models (Larson \& Tinsley 1978; van den Hoek et al.
1997) than the maximum disk values.  The need for an extra dark
component in the disk has  disappeared.

\subsection{Bottema disk halo parameters}

Whereas in the maximum disk case the galaxies with the highest values
of $V_{\rm max}$ have the lowest central halo densities, in the
Bottema disk case they have the densest and most compact halos.
The central densities of the galaxies with low $V_{\rm max}$ are, as
expected, less affected. The large change in halo parameters of HSB
galaxies compared to their maximum disk values, results in trends of
halo density and core radius opposite to those derived for maximum
disk.

Paradoxically, the uncertainties in the decompositions of the {\it
HSB} galaxies makes it very hard to make definite statements on the
halo properties and their possible changes with surface brightness.
 
\subsection{Halo parameters and galaxy evolution} 

Generalizing (and running the risk of over-interpreting) the maximum
disk and Bottema disk decompositions leads to two distinct different
sets of relations between the dark and visible components of galaxies.
The maximum disk decompositions suggest that at fixed $V_{\rm max}$
all halos are equal, while the Bottema disk decompositions suggest
that the sizes and densities of DM halos are directly related to those
of the optical disks that inhabit these halos.  In summary:
\begin{enumerate}
\item {\bf Maximum disk} Towards lower surface brightnesses halos become
denser and smaller. Disks of lower surface brightness  have higher $(M/L)_{\star}$ at fixed $V_{\rm
max}$, and extend further out into the halo.
\item{\bf Bottema disk} Towards lower surface brightnesses halos become
more diffuse and more extended. Disks of lower surface brightness have similar $(M/L)_{\star}$ at
fixed $V_{\rm max}$ and are found in more massive halos.
\end{enumerate}

Option (i) implies that the dimmest and most extended (LSB) galaxies 
live in the most compact halos.  Presumably the collapse of their massive,
dark disks was not as complete as in HSB galaxies (because of
the larger angular momenta of LSB galaxies?).  

With option (ii) the properties of the baryonic part of the galaxy are
reflected in those of the host halo: dim, diffuse and extended
galaxies inhabit extended, low density halos.  In both cosmologies the
dominance of DM increases towards lower surface brightnesses.

\subsubsection{Maximum disk cosmology}

If, for the moment, we assume the maximum disk results to hold, what
are the implications for the evolution of LSB galaxies? In other
words, what if the only difference between HSB and LSB galaxies at
fixed $V_{\rm max}$ is that LSB galaxies stick out further into
otherwise identical halos?

The picture which can then be sketched is deceivingly simple.  For a
HSB and a LSB galaxy at identical positions on the TF relation, the
LSB galaxy has to be more extended optically (Zwaan et al.  1995), and
in order to have the same luminosity as the HSB galaxy, the amount of
past and present star formation per area (the surface brightness) has
to be lower\footnote{The bluer colors of LSB galaxies are explained by
  metallicity effects and the less prominent old population. An equal
  amount of star formation has a much larger effect on the colors of a
  LSB galaxy than on a HSB galaxy (de Blok et al.  1995).}.  The LSB
optical disk simply encompasses more DM than the HSB optical disk, due
to its larger extent in an otherwise identical halo, resulting in a
LSB optically extended galaxy with a high total (i.e.  DM and luminous
matter) $M/L$.  The low star formation rate (as averaged over the
lifetime of the galaxy) implies slow evolution in the LSB galaxy and a
low $(M/L)_{\star}$ \footnote{Here we specifically do not include
  H{\sc i} and other potentially star forming material; the low
  $(M/L)_{\star}$ simply reflects that the LSB has not had enough time
  to build a large old stellar population as in HSB galaxies.}, which
is difficult to reconcile with the maximum disk hypothesis assumed to
hold to derive the these conclusions.

Current ideas on the formation and evolution of LSB galaxies (Bothun
et al.  1993, McGaugh 1992, Mo et al 1994, Dalcanton et al. 1997)
suggest that LSB galaxies form from lower amplitude density peaks than
HSB galaxies.  As they are also found to be more isolated from their
neighbors, and to have a lack of companions with respect to HSB
galaxies, they have presumably also undergone less interaction with
other galaxies.  The extended gas disks also suggest that the collapse
of the baryonic matter has been much less efficient than in actively
star forming HSB galaxies.

Although the Bottema disk solution may seems the most one, both
solutions do have their problems.  The maximum disk solution has to
explain why galaxies in otherwise identical halos have different
collapse-histories as a function of surface brightness, while the
other solution has to explain how galaxies with different halo masses
but identical luminosities conspire to still end up at the same
position on Tully-Fisher relation.

One conclusion is firm however, the dynamics of LSB galaxies are
fundamentally different from those of HSB galaxies. For example, the
rotation curves of LSB galaxies can not be explained in $\Omega_0 =1$
cosmological simulations (Moore 1994, Navarro 1997), which is a
problem, certainly in view of the implied numerical richness of LSB
galaxies (McGaugh 1995).

\section{Mass ratios}

\subsection{Importance of atomic gas}

\begin{figure*}
\epsfxsize=9cm
\hfil\epsfbox{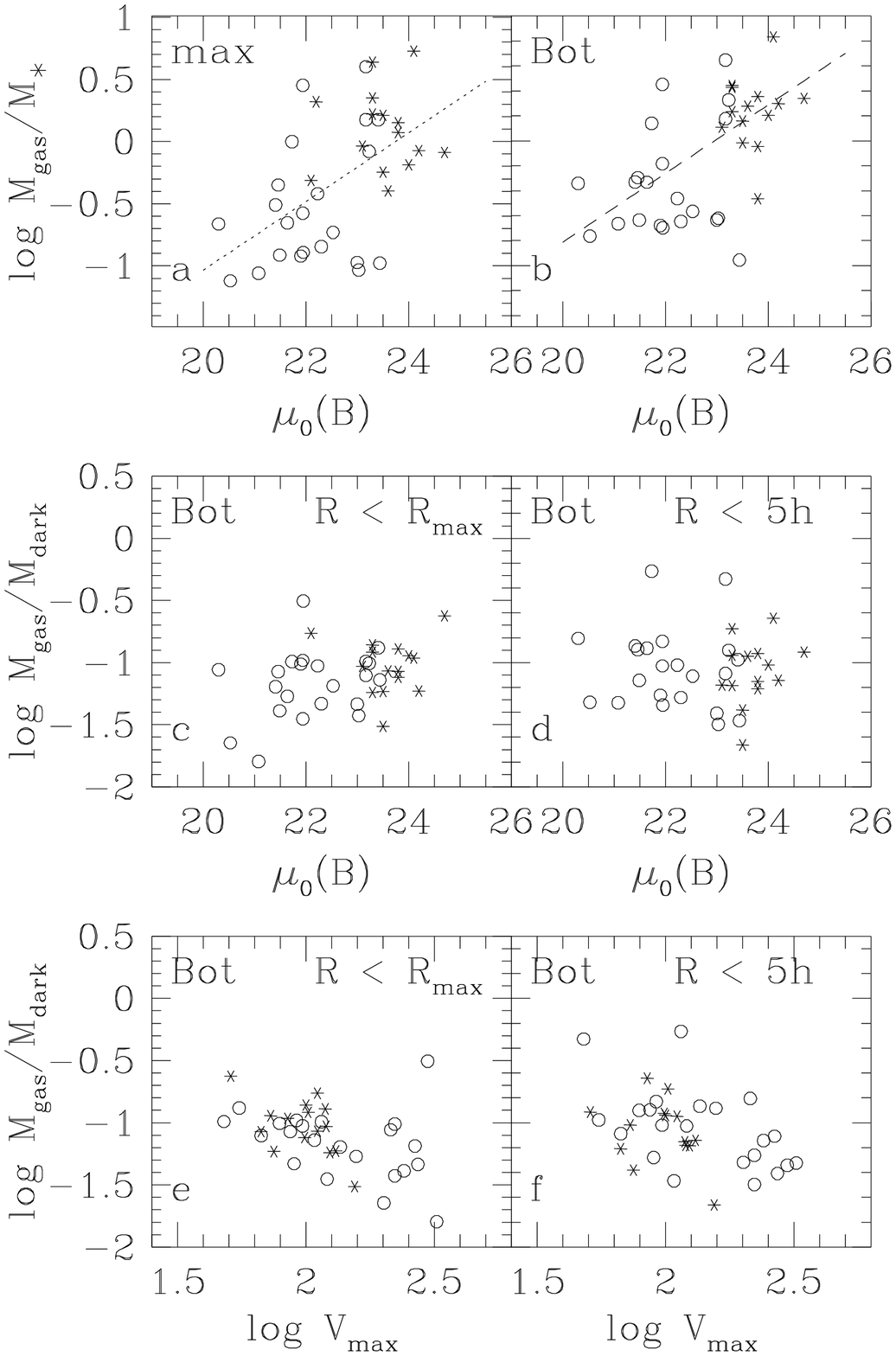}\epsfxsize=9cm\epsfbox{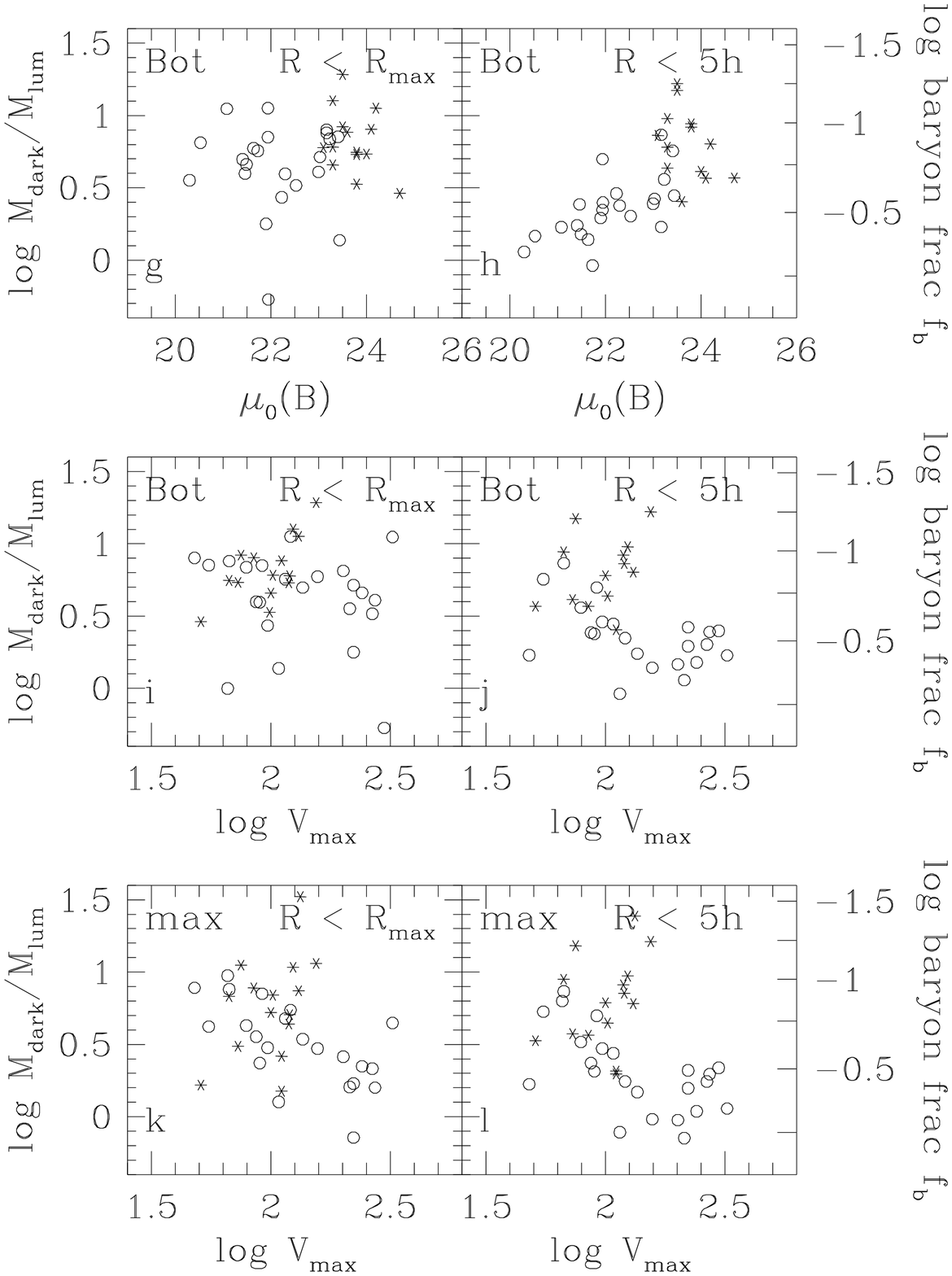}\hfil
\caption{Ratios of the various mass components. All stellar and dark
  masses are computed as indicated in the panels, where `max' denotes
  maximum disk and `Bot' Bottema disk.  The outermost radius $R_{\rm
    max}$ is defined as the the radius of the last measured point of
  the rotation curve. $h$ is the disk scale length.  Shown are: the
  ratio of stellar masses and gas masses for maximum disk ({\bf a}),
  where the dotted line is a least-squares fit to the data; the ratio
  of stellar masses and gas masses for Bottema disk ({\bf b}); the
  ratio of gas mass and dark mass as a function of surface brightness
  $\mu_0$, measured within $R_{\rm max}$ ({\bf c}) and within 5$h$
  ({\bf d}); the ratio of gas mass and dark mass as a function of
  maximum rotation velocity $V_{\rm max}$, measured within $R_{\rm
    max}$ ({\bf e}) and within 5$h$ ({\bf f}); the ratio of dark mass
  and luminous mass (and baryon fraction) as a function of $\mu_0$,
  measured within $R_{\rm max}$ ({\bf g}) and within 5$h$ ({\bf h});
  the ratio of dark mass and luminous mass for the Bottema disk as a
  function of $V_{\rm max}$, measured within $R_{\rm max}$ ({\bf i})
  and within 5$h$ ({\bf j}); the ratio of dark mass and luminous mass
  for the maximum disk as a function of $V_{\rm max}$, measured within
  $R_{\rm max}$ ({\bf k}) and within 5$h$ ({\bf l})}
\label{Mastot}
\end{figure*}

The increasing dynamical importance of the gas component can readily be
seen in Fig \ref{Mastot}a and b.  Galaxies of lower surface brightnesses
have larger values of $M_{\rm gas}/M_{\star}$.  A similar trend with
Hubble type was already noticed by e.g.\ Broeils (1992). 
The trend is clear in both the maximum disk and Bottema cases,
although of course the gas-to-stellar mass ratios derived using
maximum disk stellar masses tend to be larger.  The ratio ranges from
$\sim 0.2$ for the HSB  galaxies, to $\sim 2$ in
the LSB galaxies.  Clearly in LSB galaxies the gas component can be of
a larger dynamical importance than the stellar component, even in the
maximum disk case.  This effect is more pronounced for the Bottema
decompositions because of the limited range in $(M/L)_{\star}$.  A
fuller discussion of the gas fraction over a large range in surface
brightness is presented in McGaugh \& de Blok (1997).

$M_{\rm gas}/M_{\rm dark}$ (as computed within the outermost measured
point of the rotation curve, $R_{\rm max}$ (see Table 1), and
assuming Bottema disk) is shown in Fig.  \ref{Mastot}c and e.  Panels
d and f show $M_{\rm gas}/M_{\rm dark}$ measured within 5$h$.
There is clear trend of increasing gas-richness towards low values of
$V_{\rm max}$, an effect which is not entirely cancelled when measured
within 5$h$. Slower rotating galaxies contain relatively more gas. The
large range in $\mu_0$ found at each $V_{\rm max}$ ensures that no
clear trend is visible of $M_{\rm gas}/M_{\rm dark}$ with $\mu_0$.

\subsection{Baryon fraction and dark matter fraction}

The total luminous mass $M_{\rm lum}$ is defined as the sum of stellar
mass and the gas mass. We here take the stellar mass as implied by the
Bottema disk. The ratio of the dark and luminous mass is
related to the maximum velocity (or luminosity) (Persic \& Salluci
1991, Broeils 1992).  Galaxies with low maximum velocities and low
luminosities are more DM dominated than high luminosity galaxies.
This effect is not very pronounced when $M_{\rm dark}$ is computed
within the outermost radius (Fig. \ref{Mastot}g and i): the extended
and well-observed rotation curves of HSB galaxies encompass a
relatively much larger fraction of the halo.  The effect becomes much
more pronounced when measured within a fixed number of scale lengths
(Fig.  \ref{Mastot}h and j). 

The extreme DM domination of the lowest surface brightness
galaxies has more general implications.  One example is the
baryon fraction $f_b$ of the universe.  This is of direct cosmological
significance since $\Omega_0 = \Omega_b/f_b$, where $\Omega_b$ is the
baryon density of the universe: a measure of $f_b$
combined with $\Omega_b$ from primordial nucleosynthesis directly
yields $\Omega_0$.

A value of $f_b$ thought to be representative of the universal baryon
fraction is given by X-ray observations of galaxy clusters (White et
al.\ 1993; David et al.\ 1995; White \& Fabian 1995): $0.1 < f_b <
0.2$.  A similar result is obtained for compact groups of galaxies
(Pildis et al.\ 1995).  Primordial nucleosynthesis gives $\Omega_b =
0.022$ (e.g.\ Walker et al.\ 1991); together these suggest
$\Omega_0 \approx 0.2$.

Now consider the rotation curve data.  Dark matter is required to keep
the rotation curves flat, and since there is no indication of
declining rotation curves, the halos must be substantially larger than
the last measured point.  The dark to luminous mass within $R_{\rm max}$
therefore gives an upper limit on $f_b$ (Fig.\ \ref{Mastot}), through
$f_b < (M_{\rm lum}/M_{\rm tot})$ and therefore
\[ f_b < \left( 1+ {M_{\rm dark}\over{M_{\rm lum}}} \right)^{-1}.\]

To be conservative, let us consider the maximum disk case.  For most
HSB galaxies, $R_{\rm max}$ is not sufficiently large to provide an
interesting limit.  However, the DM domination of LSB galaxies
improves this situation.  In our sample, there are a few galaxies for
which the upper limit on $f_b$ is lower than the lower bound of the
range of the cluster baryon fractions.  These are F568--V1, UGC 5750,
UGC 5999 (all with $f_b < 0.08$), and F571--8 ($f_b < 0.05$) (although
this latter value is uncertain due to the inclination of the galaxy).
The rotation curves of all four of these galaxies are flat or rising
at the last measured point, so the dark halos must extend well beyond
$R_{\rm max}$ to avoid having a perceptible effect.  The baryon
fractions of these galaxies really are different from those derived
for clusters of galaxies.

It is not obvious why the baryon fraction should differ from halo to
halo.  If for the moment we assume our limit to be representative of
the universal fraction, then the limits on $\Omega_b$ and $f_b$ give
$\Omega_0 > 0.45$.  However, the methods used to obtain the cluster
results have now been thoroughly analyzed by many groups (e.g.\ White
\& Fabian 1995; Evrard et al.\ 1996) and seem robust.

It may be that $f_b$ really varies from halo to halo, perhaps
depending on scale.  What physics might cause this is unclear.
Dissipation acts in the wrong direction: if galaxies have dissipated a
lot, their halos are very large, worsening the problem.  It has been
suggested that low mass galaxies experience significant wind driven
mass loss (e.g.\ Dekel \& Silk 1986) which might rid them of some
baryons.  However, this clearly has not occurred in the galaxies we
are discussing.  Their gas has not been swept out; they are very gas
rich and massive.  They show no evidence of blow-out events (Bothun et
al.\ 1994).  It is therefore entirely ad-hoc to vary $f_b$, and this
would add yet another free parameter to the current cosmological
models.

Of course, until the nature of the dark matter is known, it is
conceivable that some or all of it is baryonic.  Thus, another
possibility is that some fraction of halos is baryonic, and relatively
more baryons have been incorporated into the halos of LSB galaxies,
perhaps as MACHOs.  How and why this should occur is again inobvious.
LSB galaxies show little evidence of an early epoch of star formation
which ought to provide the remnants that become MACHOs (McGaugh \&
Bothun 1994; de Blok et al.\ 1995), and it is not obvious that these
should be distributed in a halo.  Completely non-baryonic matter is
still required unless primordial nucleosynthesis or large scale
estimates of $\Omega_0$ are wrong.  The limits derived from the LSB
galaxies are, however, hard limits, and they should be reconciled with
the cluster values. If that does not prove to be possible, this might
at last resort even indicate the break down of the dark matter
hypothesis (Sanders 1997).

\section{Surface and volume densities}

\subsection{Total mass volume densities}

\begin{figure}
\epsfxsize=8.5cm
\hfil\epsfbox{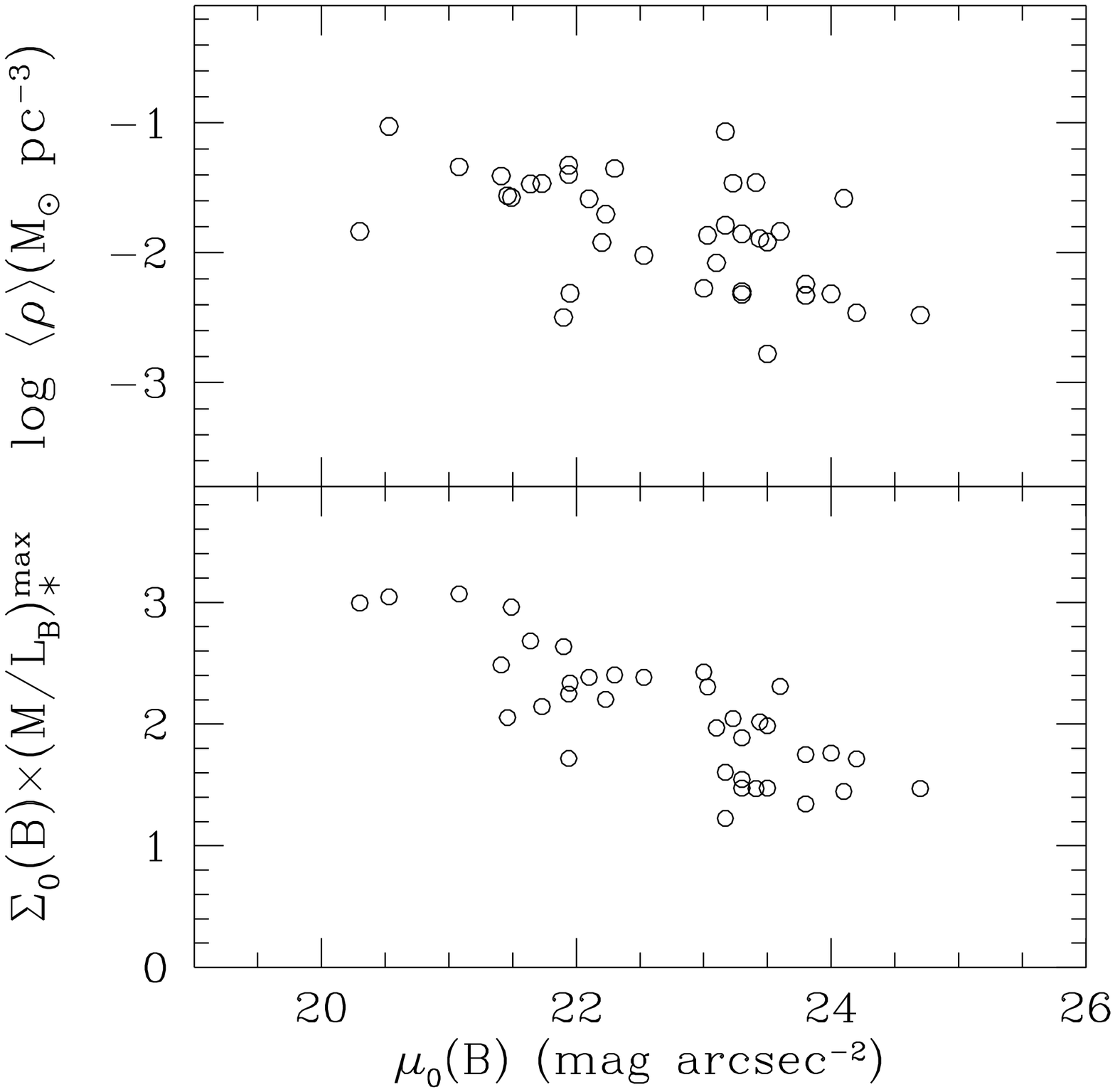}\hfil
\caption{Top panel: Average volume density of LSB galaxies as measured
  within 5 disk scale lengths. Bottom panel: Central mass surface
  density implied by the maximum disk fits. Both quantities depend on
  surface brightness.}
\label{voldens}
\end{figure}

A question which remains to be answered is whether the
low surface densities now observed in LSB galaxies also imply that they
are truly low {\it volume} density objects.  This question is
extensively discussed in de Blok \& McGaugh (1996) where the LSB galaxy
UGC 128 and HSB galaxy NGC 2403 (which lie at identical positions on the
TF relation) are compared.  The conclusion reached there is that the
total mass volume density enclosed within a fixed number of scale
lengths is a factor $\sim 10$ lower in UGC 128 than in NGC 2403.  We
refer to that work for a discussion on the preferred length scale for
measuring densities in galaxies. 

We have extended this analysis to our samples of HSB and LSB galaxies.
The total mass density within $5h$ was computed and shown as a
function of surface brightness in Fig \ref{voldens}.  The conclusion
reached in de Blok \& McGaugh is confirmed: the difference in $\langle
\rho \rangle$ between Freeman like galaxies and LSB galaxies with $\mu
\simeq 24$ is again a factor of 10.

Although a large range in mass density is inferred at each value of
$\mu_0$, LSB galaxies are ``diffuse'' objects, even in the maximum
disk case.  The increase in optical scale length towards lower surface
brightnesses is much faster than any change in the DM length scale.
The increase in area spanned by the optical disk can therefore not be
compensated by a corresponding increase in total matter enclosed by
the optical disk.  The enclosed mass increases with $V^2 h$, while the
enclosed area (volume) increases with $h^2$ ($h^3$).  The increasing
scale length therefore ensures that for LSB galaxies the average
volume or surface density enclosed by the disk is always lower.

\subsection{Maximum disk baryonic surface densities}

The maximum disk results are a good way to demonstrate that LSB
galaxies have truly low matter surface density disks.  That is, they
are not normal surface density disks that happen to have less light
per square parsec, but the underlying mass surface density
distribution is more diffuse too (de Blok \& McGaugh 1996).

In Fig \ref{voldens} we show the central surface densities calculated
by multiplying the central surface brightness with the respective
$(M/L)_{\star}$ ratios. The resulting surface density is the maximum
density the disk can have within the constraints of the observed
rotation curve.  These values have not been corrected for the presence
of H{\sc i}, as the surface density of the \HI gas becomes progressively
lower towards lower surface brightnesses (BMH96).  The increase in
maximum disk $(M/L)_{\star}$ ratio is not fast enough to compensate
for the decrease in surface brightness: an effect which will only
become more pronounced when hypotheses other than maximum disk are
used.  Maximum disk is the most conservative scenario, but already
here it is clear that the mass surface density in the disk is a
function of surface brightness.

\section{Summary}

We have presented rotation curve decompositions of a sample of LSB
galaxies and compared the properties of their constituent mass
components with those of a sample of HSB galaxies. 

LSB galaxies are very much dominated by dark matter thus proving to be
the ideal test-cases for theories on the structure and formation of
dark matter halos. The halo parameters derived for these galaxies are
insensitive to the assumed mass-to-light ratio of the stellar disk,
ensuring that the value we derive are in all probability close to the
true values.

This is in sharp contrast with HSB galaxies, where the halo parameters
are sensitive to the assumed stellar mass-to-light ratio.  Comparison
of the HSB and LSB samples shows that the maximum disk solutions imply
an increasing stellar mass-to-light ratio for the stellar disk and an
increasing compactness for the halo as the central surface brightness
decreases, while the Bottema disk solutions imply exactly the
opposite. We argue, based on what is known about the evolutionary
state of LSB galaxies from other sources, that the maximum disk
solution is not representative.  To explain the maximum disk results
in a physical way, one has to assume large amounts of baryonic (dark)
matter in the LSB disks.  This would result in massive, evolved disks,
which is not consistent with the observed past and present low star
formation rates of LSB galaxies.  The Bottema disk solutions give
lower $(M/L)_{\star}$ values, more consistent with what can be derived
from other evolutionary probes (metallicity, gas fraction).  They imply
that LSB galaxies live in extended diffuse halos.  The halos of LSB
galaxies are thus fundamentally different from those of HSB galaxies
of the same mass.  This is a challenge for cosmological N-body
simulations which will  have to address this problem,
especially considering the numerical richness of LSB galaxies.

LSB galaxies are thus slowly evolving, isolated, diffuse, low-density
and extremely dark matter dominated galaxies.
 
\section*{Acknowledgments}
We thank Roelof Bottema for allowing us to use his results prior to
publication and for his extensive comments on early drafts of this
paper.  We also would like to thank Thijs van der Hulst, Renzo Sancisi
and Marc Verheijen for the many discussions.  We thank Floor Sicking
and Kor Begeman for allowing us to use their rotation curve fitting
software. We  acknowledge remarks made by Dr.\ Kalnajs and the
anonymous referee which helped to clarify some aspects of this paper.

\appendix
 \section{The generalized Bottema-disk}

 \subsection{Introduction}

 Bottema (1995) has shown from measurements of the stellar velocity
 dispersions in a sample of HSB galaxies that the maximum rotation
 velocity of the stellar disk (i.e.  the velocity occuring at 2.2$h$)
 is on average equal to 63\% of the measured rotation velocity at that
 radius.

 This result was derived under the explicit assumption of Freeman's
 Law (which applies to most of the galaxies in Bottema's sample).  It
 is therefore not possible to directly apply Bottema's result to our
 sample of LSB galaxies, especially if one realizes that 63\% of the
 observed rotation velocity in many LSB galaxies exceeds the maximum
 disk solution.

 In Bottema (1997) a re-analysis and generalization of these results
 is presented, also taking into account surface brightness and color
 effects. We here summarize the main points of that analysis.

\subsubsection{Poor man's population synthesis}

The disk is assumed to be unaffected by dust and
contains only two populations (van der Kruit 1986; Bottema 1988): a
young component with a color $(B-V)^y = -0.03$ and an old component
with $(B-V)^o = 0.97$. The old component furthermore contains all of
the disk mass.

It is straightforward to show that this implies that $S_V^y = 0.973
S_B^y$ and $S_V^o = 2.44 S_B^o$, where $S_{B,V}^{o,y}$ are the fluxes
of the old and young populations in the $B$ and $V$ bands.

The absolute magnitude of the old disk in the $B$-band, $M_B^{od}$, 
is related to the total absolute magnitude $M_B^{tot}$ by
\begin{equation}
M^{\rm od}_B = M_B^{\rm tot} - 2.5 \log (S_B^y / S_B^{\rm tot})
\end{equation}

It is straightforward to show that this equation and the colors of
the populations imply that 

\begin{equation} M_{\rm od}^B = M_B - 2.5 \log \left( {{1-0.973 \cdot A^{\star}} \over
{1.467 \cdot A^{\star} }}\right) = M_B - cf,\end{equation}
where $A^{\star}$ is defined as $10^{-0.4(B-V)}$.

\subsubsection{The exponential disk}

For an exponential, locally isothermal, constant $M/L$ disk with
density distribution $\rho(R,z)=\rho(R,0)\,{\rm sech}^2(z/z_0)$, 
and \begin{equation} \langle v_z^2 \rangle^{\frac{1}{2}}_{R=0} =
  \sqrt{\pi G \sigma_0 z_0}, \end{equation} and
\begin{equation}\langle v_z^2 \rangle^{\frac{1}{2}}_{R=0} = \sqrt{\pi G\, \Sigma_0
    \left({{M}\over{L}}\right) z_0}, \end{equation} where $\langle v_z^2
  \rangle^{\frac{1}{2}}_{R=0}$ is the central velocity dispersion, and
  $\sigma_0$, $\Sigma_0$ and $z_0$ are the central surface density,
  surface brightness and scale height, respectively.

For an infinitely thin stellar disk the maximum rotation velocity is
given by \begin{equation}v_{\rm max} = 0.88 \sqrt{\pi G\, \sigma_0 h}.\end{equation}
Combining Eqs. (A4) and (A5) yields
\begin{equation}v_{\rm max} = 0.88\, \langle v_z^2 \rangle^{\frac{1}{2}}_{R=0} \sqrt{{h}\over{z_0}}.\end{equation}

\subsubsection{The old-disk Tully-Fisher relation}

Equation (A5) and the identity $L=2\pi\Sigma_0 h^2$,
where $L$ is the total luminosity, $\Sigma_0$
the central surface brightness in linear units, and $h$ the scale
length, imply an old disk Tully-Fisher relation
\begin{equation}
v^4_{\rm max, od} = 0.3 \pi G^2 \Sigma_0^{\rm od} \left( {{M}\over{L}}
\right)^2_{\rm od} L_{\rm od}
\end{equation}
As the old disk contains all the mass $v^4_{\rm max,od} = v^4_{\rm max}$.
$(M/L)_{\rm od}$ is assumed to be the the same for all old disks.

\subsection{Bottema disk for Freeman galaxies}

For Freeman galaxies $\mu_0$ is constant. The color is also assumed
to be constant at $B-V = 0.7$.
Eq. (A7) then yields
$ M_{\rm od} = 
-10\, \log (v_{\rm max}) + C$, or 
\begin{equation} v_{\rm max} = 10^{0.1 C} 10^{-0.1 M_{\rm od}}. \end{equation}
Combining Eqs. (A6) and (A8) yields
\begin{equation}\langle v_z^2 \rangle^{\frac{1}{2}}_{R=0} = A^{-1}
  \sqrt{{z_0}\over{h}}\, 10^{-0.1 M_{\rm od}},\end{equation}
where $A$ is defined as \begin{equation}A=0.88 \cdot 10^{-0.1 C}.\end{equation}
Following Bottema (1995) the empirical relation between
luminosity and scale length ratios is adopted
$h/z_0 = 0.6\, M_{\rm od} + 17.5$ together with  the corresponding values
$A = 0.75$, $C= 0.69$ and $(M/L_B) = 1.85$. This results in

\begin{equation}v_{\rm max} = 1.17 \cdot 10^{-0.1 M_{\rm od}}.\end{equation}

\subsection{General Bottema disk for non-Freeman galaxies}

The result derived above strictly speaking only applies to galaxies that obey
Freeman's Law. To generalize to different surface brightnesses 
rewrite Equation (A7) as
\begin{equation}v^4_{\rm max} = 0.3 \pi G^2
  \Sigma_{0,F}\left({{\Sigma_0^{\rm od}}\over{\Sigma_{0,F}}}\right) \left( {{M}\over{L}}
\right)^2_{\rm od} L_{\rm od},
\end{equation}
where $\Sigma_{0,F}$ is the constant Freeman central surface brightness
of 136 $L_{\odot}$ pc$^{-2}$. The scale lengths of
the old and young populations are assumed to be equal, so that
$\Sigma_0^{\rm od}/\Sigma_0 = L_{\rm od}/L_{\rm tot} = 10^{0.4cf}$.
This results in
\begin{equation}
v_{max} = {\rm const} \left( {{\Sigma_0}\over{\Sigma_{0,F}}}
\right)^{\frac{1}{4}} 10^{0.1cf}10^{-0.1M_{\rm od}}
\end{equation}

For Bottema's sample, with central surface brightnesses approximately
equal to the Freeman value and colors close to $B-V = 0.7$, 
an average correction factor $\langle cf \rangle \simeq -0.5$ can be defined.
Combination of Eqs (A6) and (A13) then yields
\begin{equation}
  \langle v_z^2 \rangle^{\frac{1}{2}}_{R=0} = {{\rm const}\over{0.88}}
  \sqrt{{z_0}\over{h}} \left(
    {{\Sigma_0}\over{\Sigma_{0,F}}} \right)^{\frac{1}{4}} 
 {{{10^{0.1(cf+\langle cf \rangle-M_{\rm od})}}}\over{10^{0.1\langle cf \rangle}}}
\end{equation}

Again using Bottema's relation $h/z_0 = 0.6M_{\rm od}+17.5$ then implies
that 
\[ {\rm const} \times 10^{0.1\langle cf \rangle} = 1.17 \]

For the maximum rotation velocity and velocity dispersion of a
galactic disk one then finds (using Eqs. (A13) and (A14))
\begin{equation}v_{\rm max} = 1.17 \left({{\Sigma_0}\over{\Sigma_{0,F}}}
  \right)^{\frac{1}{4}}\cdot 10^{0.1(cf - \langle cf \rangle)}
  10^{-0.1 M_{\rm od}},\end{equation} and
\begin{equation}
  \langle v_z^2 \rangle^{\frac{1}{2}}_{R=0} = 1.33
  \sqrt{{z_0}\over{h}} \left(
    {{\Sigma_0}\over{\Sigma_{0,F}}} \right)^{\frac{1}{4}} 
 {{10^{0.1(cf-\langle cf \rangle-M_{\rm od})}}}
\end{equation}

Conversion to observables using Eq (A2) then yields
\begin{equation}v_{\rm max} = 1.17 \left({{\Sigma_0}\over{\Sigma_{0,F}}}
  \right)^{\frac{1}{4}}\cdot 10^{0.2(cf +0.25)}10^{-0.1 M_B^{\rm tot}},\end{equation}

\subsection{Mass-to-light ratio}
From Eqs. (A4) and (A16) one can derive that

\begin{equation} 
\left( {{M}\over{L}}\right)_B = 28.1 \cdot 10^{-0.2
    M_{B,\odot}} \cdot 10^{0.4cf}
10^{-0.2 \langle cf \rangle},
\end{equation} 
With $M_{B,\odot} = 5.48$ one finds
\begin{eqnarray} \nonumber
\left( {{M}\over{L}}\right)_B &= &1.79
   \cdot 10^{0.4cf}
10^{0.2(0.5- \langle cf \rangle)} = 2.84\cdot 10^{0.4cf}\\
& = & 1.936 \cdot 10^{0.4(B-V)} - 1.883.
\end{eqnarray} 
The mass-to-light ratio does not depend on the surface brightness of
the disk, which is due to our assumption that the old disk population
has the same mass-to-light ratio for all disks. Disks with identical
colors therefore have identical mass-to-light ratios, irrespective of
surface brightness.

\end{document}